\documentclass[sigconf,nonacm]{acmart}
\AtBeginDocument{%
  }

\setcopyright{none}

\usepackage{booktabs}
\usepackage{multirow}
\usepackage{tikz}
\usepackage{subcaption}
\usepackage{makecell}
\usepackage{colortbl}
\usepackage{array}

\usepackage[linesnumbered,ruled,vlined]{algorithm2e}

\newcommand{\ourtool}{CryptoGuard}

\newcommand*\fullcirc{
\begin{tikzpicture}[baseline=-0.6ex]
    \fill (0,0) circle (0.8ex);
\end{tikzpicture}
}

\newcommand*\emptycirc{
\begin{tikzpicture}[baseline=-0.6ex]
    \draw[black,fill=white] (0,0) circle (0.8ex);
\end{tikzpicture}
}

\begin{document}

\title{CryptoGuard: Lightweight Hybrid Detection and Response to Host-based Cryptojackers in Linux Cloud Environments}
\thanks{\textit{This is the author's accepted manuscript of the article published in the 
Proceedings of the 20th ACM Asia Conference on Computer and Communications Security (ASIACCS '25), 
August 25–29, 2025, Hanoi, Vietnam. 
DOI: \url{https://doi.org/10.1145/3708821.3736186}.}}

\author{Gyeonghoon Park}
\email{hoonp2@uci.edu}
\orcid{0009-0006-1085-8774}
\affiliation{%
  \institution{UC Irvine}
  \city{Irvine}
  \state{CA}
  \country{USA}
}

\author{Jaehan Kim}
\email{jaehan@kaist.ac.kr}
\orcid{0000-0001-8048-097X}
\affiliation{%
  \institution{KAIST}
  \city{Daejeon}
  \country{Republic of Korea}
}

\author{Jinu Choi}
\email{wlsgudy3@kw.ac.kr}
\orcid{0009-0004-1430-757X}
\affiliation{%
  \institution{Kwangwoon University}
  \city{Seoul}
  \country{Republic of Korea}
}

\author{Jinwoo Kim}
\authornote{Corresponding author}
\email{jinwookim@kw.ac.kr}
\orcid{0000-0003-1303-8668}
\affiliation{%
 \institution{Kwangwoon University}
 \city{Seoul}
 \country{Republic of Korea}}

\renewcommand{\shortauthors}{Park et al.}

\begin{abstract}
Host-based cryptomining malware, commonly known as cryptojackers, have gained notoriety for their stealth and the significant financial losses they cause in Linux-based cloud environments. Existing solutions often struggle with scalability due to high monitoring overhead, low detection accuracy against obfuscated behavior, and lack of integrated remediation. We present CryptoGuard, a lightweight hybrid solution that combines detection and remediation strategies to counter cryptojackers. To ensure scalability, CryptoGuard uses sketch- and sliding window-based syscall monitoring to collect behavior patterns with minimal overhead. It decomposes the classification task into a two-phase process, leveraging deep learning models to identify suspicious activity with high precision. To counter evasion techniques such as entry point poisoning and PID manipulation, CryptoGuard integrates targeted remediation mechanisms based on eBPF, a modern Linux kernel feature deployable on any compatible host. Evaluated on 123 real-world cryptojacker samples, it achieves average F1-scores of 96.12\% and 92.26\% across the two phases, and outperforms state-of-the-art baselines in terms of true and false positive rates, while incurring only 0.06\% CPU overhead per host.
\end{abstract}



\maketitle

\section{Introduction}

The cybersecurity landscape has witnessed the emergence of a novel threat in the form of cryptomining malware, commonly referred to as \emph{cryptojackers}. Cryptojackers denote malicious software designed to engage in unauthorized cryptomining on victimized computers, operating without the user's knowledge or consent. Unlike traditional malware that directly targets and compromises user resources through information leaks or denial-of-service attacks, cryptojackers exploit computing power to generate illicit monetary gains. Cryptomining, considered a legitimate activity, complicates the task of distinguishing between lawful and illegal instances. Consequently, the prevalence of cryptojackers has surged, with over 300 million incidents reported in 2023, according to the SonicWall security report~\cite{sonicwall}. A notable incident involved Tesla, where attackers mined more than \$3 million through Monero cryptojackers running in public cloud environments~\cite{tesla}.

Cryptojackers are typically categorized into two main types: (i) \emph{in-browser} and (ii) \emph{host-based}~\cite{tekiner2021sok}. In-browser cryptojackers are well-known for exploiting client web browsers connected to malicious websites, resulting in numerous victims. In contrast, host-based cryptojackers have received comparatively less attention due to their need to compromise hosts or servers. However, the growing adoption of cloud computing has led to a significant increase in Linux-based hosts~\cite{cloud90linux}, such as virtual machines and containers, which has, in turn, facilitated the rise of Linux-based cryptojackers~\cite{vmware_report, docker}. A prominent example is an exploit known as GUI-vil, which targets Amazon EC2 instances and is attributed to Indonesian cybercriminals~\cite{indonesian}. Given these escalating threats, it is imperative to develop effective defenses against host-based cryptojackers.

Existing literature proposes numerous solutions focused on either (i) \emph{detecting} cryptojackers or (ii) \emph{remediating} hosts compromised by cryptojackers. Given the elusive nature of cryptojacker behaviors designed to evade detection for extended periods, many detection solutions~\cite{barbhuiya2018rads,ahmad2020dca,mani2020decryptopro,gangwal2020hpc,darabian2020syscall,tekiner2022ali,carprolu2021noise,almurshid2023binary,gomes2020cryingjackpot} predominantly utilize machine learning approaches to identify their stealthy activity.  
In contrast, remediation strategies aim to mitigate the impact of cryptojackers on compromised hosts through diverse countermeasures, such as filtering~\cite{franco2023honeypot} and sandboxing~\cite{sandboxing}. While the choice between detection and remediation depends on the cloud administrator's specific needs, we posit that a comprehensive approach integrating both aspects is lacking in prior work.  
Considering the prevalence of cryptojackers targeting Linux-based hosts and their persistence mechanisms exploiting common entrypoints (e.g., \texttt{cronjob}~\cite{kissadog}, \texttt{rc.local}~\cite{rc_local1,rc_local2}), it is crucial to have hybrid solutions that not only detect but also promptly remediate compromised systems to reduce the burden on cloud administrators.

However, implementing a practical hybrid solution for modern cloud environments presents several challenges. First, existing machine learning-based detection systems require the collection of fine-grained features, such as syscalls, which is difficult to achieve with minimal overhead given the limited resources of cloud hosts. Second, recent host-based cryptojackers employ obfuscation techniques, such as CPU throttling~\cite{tekiner2021sok,tekiner2022ali}, to evade threshold-based detection, making it crucial to design detection models that are resilient against such tactics. Third, host-based cryptomining often persists by exploiting Linux scheduler entrypoints or mutating PIDs. While tracing these features is essential to remediate hosts compromised by cryptojackers from continuing illegal mining activities, manually managing this task is daunting for cloud administrators due to the vast number of hosts and processes involved.

To address these challenges, this paper introduces \ourtool{}, a lightweight solution for hybrid detection and remediation of hosts compromised by cryptojackers in Linux cloud environments. Our approach utilizes deep learning models trained to recognize time-series syscall patterns, which serve as robust features for capturing cryptojackers' behavior—even under obfuscation attempts. This involves monitoring syscall activity across numerous virtual machines (VMs) and processes. To efficiently collect and analyze these patterns at scale, we decompose the monitoring task into two components: (i) sketch-based host monitoring and (ii) sliding window–based process monitoring, with each component paired with a dedicated deep learning model. While cryptojackers may obscure these patterns at the cost of operational efficiency, our models are designed to learn and detect these subtle behaviors effectively. Once cryptojackers are identified through two-phase classification, cloud administrators can implement countermeasures—such as process termination, quarantine, or network blocking—through remediation APIs provided by \ourtool{}. These APIs offer actionable interfaces for mitigating active threats introduced by cryptojackers. The solution leverages eBPF (extended Berkeley Packet Filter)~\cite{rice2023ebpf}, a Linux kernel feature that enables fast and lightweight monitoring and filtering of processes and containers.

We implemented \ourtool{} with 5,000 lines of C code, leveraging libbpf and bpftrace for its monitoring and prevention modules. For the deep learning classifiers, we employed TensorFlow to design a Convolutional Neural Network (CNN) and a Long Short-Term Memory (LSTM) network, which effectively learn time-series syscall data and outperform the traditional random forest model and existing solutions~\cite{tanana2020wmi,darabian2020syscall}. To evaluate the effectiveness and efficiency of \ourtool{}, we tested it using 123 real-world host-based cryptojacking malware samples from the VMware Threat Report 2022~\cite{vmware_samples, vmware_report}. To further demonstrate its ability to distinguish between cryptojacker and non-cryptojacker malware, we extended our evaluation with real-world Mirai~\cite{antonakakis2017understanding} malware samples obtained from MalwareBazaar~\cite{malwarebazaar}. Our results showed that the deep learning classifiers achieved F1-scores of 96.12\% and 92.26\% for their respective phases, along with 91.68\% when extended to identifying non-cryptojacker malware. Throughout testing, \ourtool{} introduced an average CPU overhead of just 0.06\% per host.

\noindent\textbf{Contributions.} Our contributions are summarized as follows:

\begin{itemize}
    \item We propose \ourtool{}, a hybrid solution for the seamless detection and remediation of hosts compromised by cryptojackers.
    \item We develop deep learning models that learn time-series syscall traces to detect cryptojackers, even when they employ obfuscation techniques.
    \item We implement the monitoring and remediation modules of \ourtool{} using eBPF, a lightweight tracing framework supported by recent Linux kernels.
    \item We evaluate the performance of \ourtool{} using 123 real-world host-based cryptojacking malware samples.
    \item To support open science and future research, we publicly release \ourtool{} and our dataset\footnote{https://github.com/PGHOON/CryptoGuard.git}.
\end{itemize}

\section{Background and Motivation}
\label{sec:motivation}

In this section, we present background knowledge for host-based cryptojackers and analyze existing defenses to motivate our work.

\subsection{Host-based Cryptojackers on Linux Cloud Environments}

Host-based cryptojackers are stealthy malware that covertly use a victim's system resources for illegal cryptomining. Unlike in-browser cryptojackers that execute web scripts remotely, host-based cryptojackers must be installed on the target system. Attackers achieve this by exploiting various vulnerabilities~\cite{vul1_news}, such as leaked credentials to infiltrate clouds~\cite{leak} or employing social engineering techniques~\cite{vul2_news} to deliver malware binaries~\cite{tekiner2021sok}. Although this approach requires more effort than in-browser cryptojacking and has thus received less attention, the rising prevalence of cloud environments has made them lucrative targets. Once attackers infiltrate virtual machines (VMs) or containers through compromised accounts~\cite{aws_news,us_cloud_news}, they can stealthily and persistently abuse cloud resources. Detecting these attacks is challenging for cloud administrators due to the sheer number of VMs or containers in operation. For example, a cryptojacker called WatchDog operated undetected for two years~\cite{watchdog_news}.

\begin{figure}[t]
    \centering
    \includegraphics[width=1\linewidth,trim={1.5cm 0 1.5cm 0}]{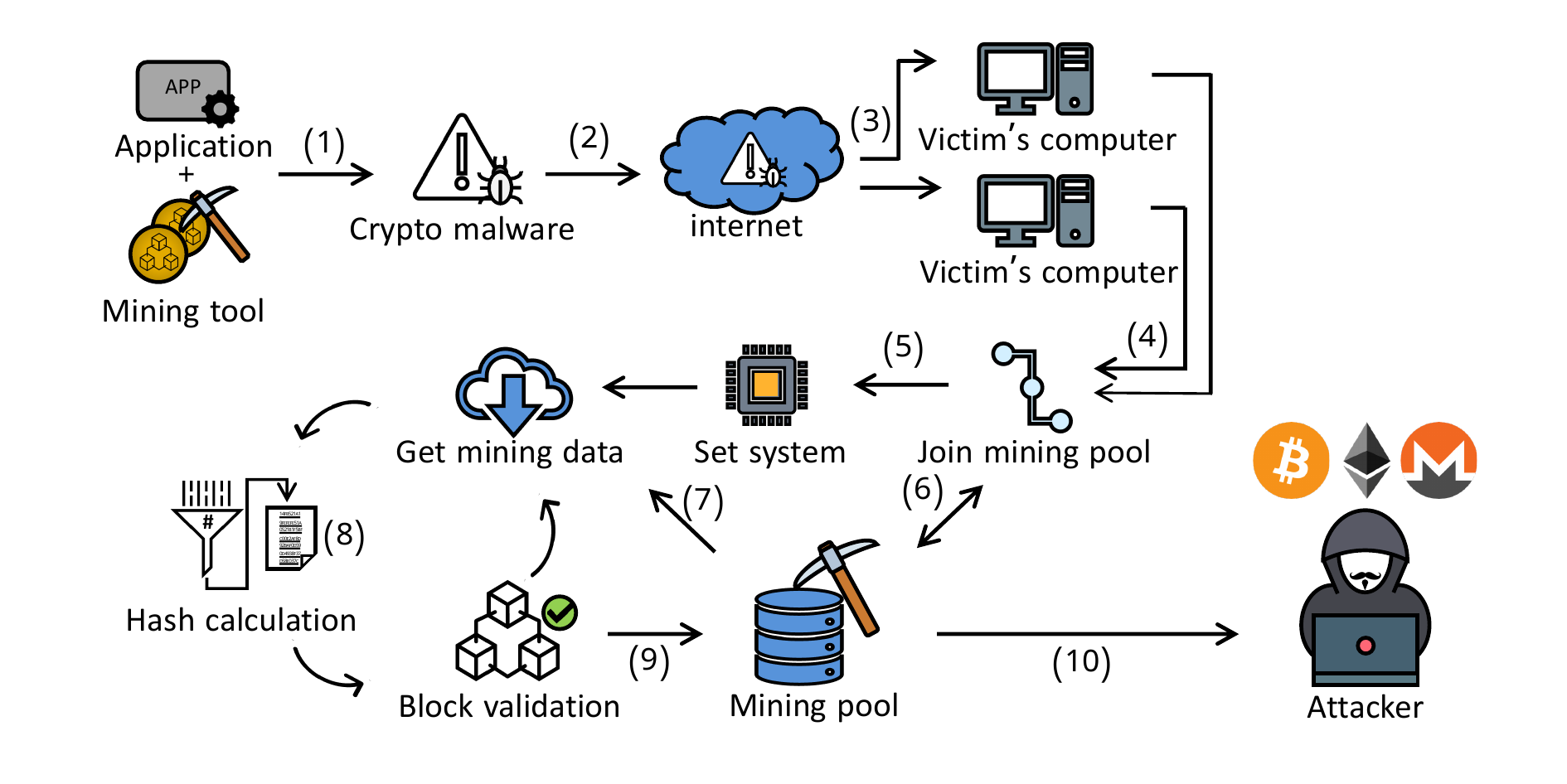}
    \caption{An illustration of the life cycle of a host-based cryptojacker.}
    \label{fig:cryptojacker}
    \vspace{-0.2in}
\end{figure}

Figure~\ref{fig:cryptojacker} shows a comprehensive overview of the lifecycle of a host-based cryptojacker~\cite{tekiner2021sok}. While its behavior varies across different variants, the typical sequence involves: (1) An attacker creates a cryptojacker by combining a mining tool (e.g., XMRig~\cite{xmrig}) with a benign application. (2) The cryptojacker is deployed to the Internet. (3) It is downloaded onto a victim's computer (or VMs, containers in a cloud environment). (4) Upon execution, the cryptojacker initializes configurations such as connecting to a mining pool with a certain protocol (e.g., Stratum~\cite{zhang2023under}) for the attacker's wallet address. (5) During its operation, the cryptojacker checks the accessibility of system resources, including CPU and memory, and allocates them accordingly. (6) The cryptojacker's mining logic runs stealthily in the background. (7) It retrieves a mining task from the pool. (8) The cryptojacker dedicates significant time and system resources to calculating a valid hash value for mining. (9) If a valid hash is found, it is transmitted to the mining pool. Following verification, the process repeats from step (7) to obtain another mining task. (10) Based on the results, the attacker achieves rewards for the completed mining process without any effort.

\subsection{Motivating Example}

\begin{figure}[t]
    \centering
    \subfloat[CPU usage patterns of benign processes and XMRig-based (\textcolor{purple}{purple} and \textcolor{pink}{pink}) and Sysrv-based (\textcolor{red}{red}) cryptojackers over time.]{
        \includegraphics[width=1\linewidth]{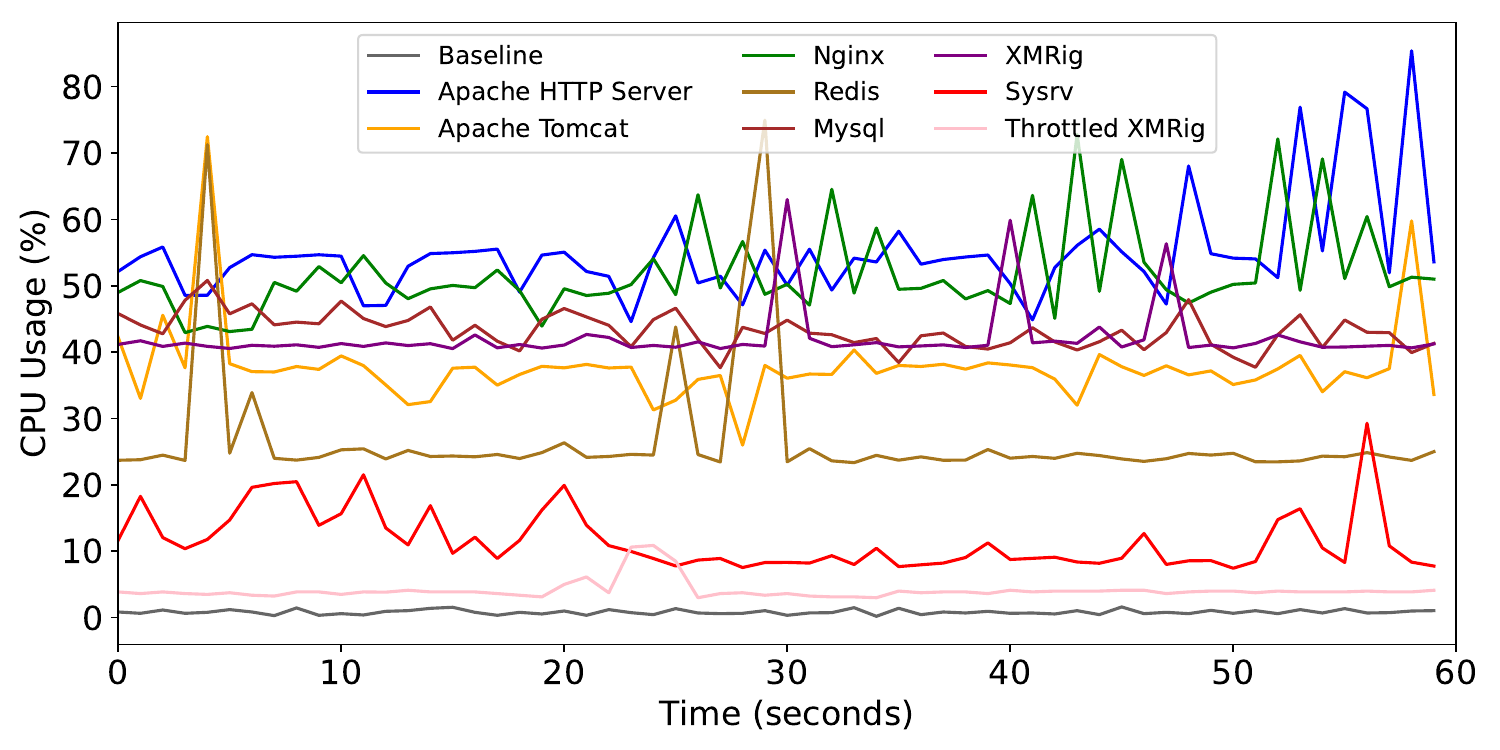}
        \label{fig:cpu_benchmark}
    
    }
    \hfill
    \subfloat[A Sysrv-based cryptojacker obfuscating its PID at different points in time.]{
    \centering
        \includegraphics[width=1\linewidth]{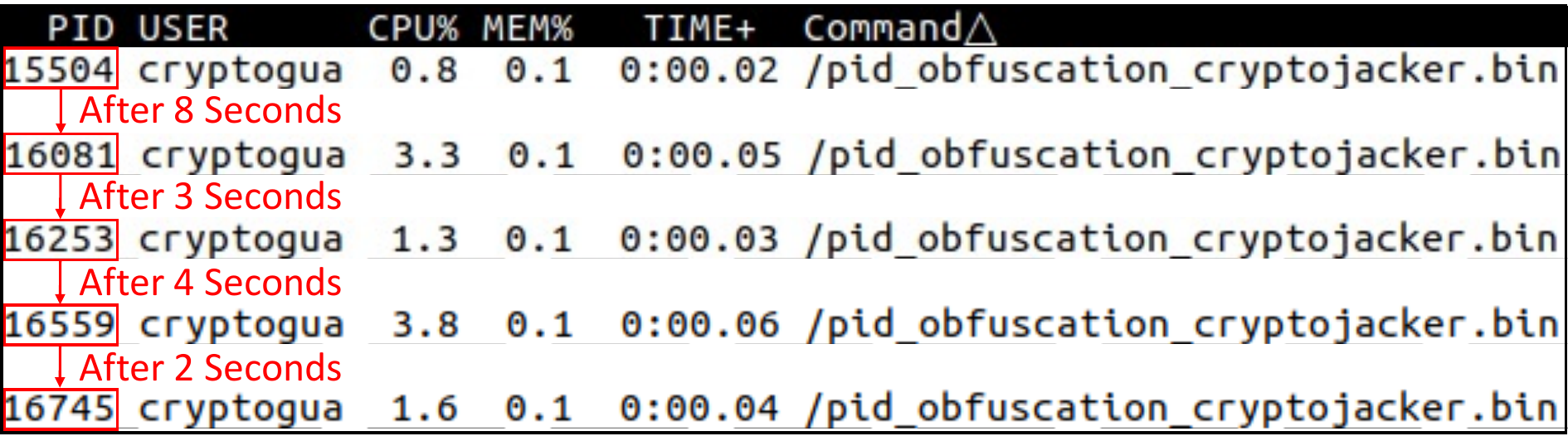}
        \label{fig:pid_obfuscation}
    }
    \caption{A motivating example.}
    \vspace{-0.2in}
\end{figure}

A naive approach to detecting host-based cryptojackers relies on rule-based detection solutions, such as CPU usage thresholds or known process signatures. However, modern host-based cryptojackers employ various \emph{obfuscation} techniques that make detection challenging in practice. To illustrate this, we conducted an experiment running benign processes (e.g., Apache HTTP Server, Apache Tomcat, Nginx, Redis, MySQL) alongside Xmrig- and Sysrv-based cryptojackers---two popular host-based cryptojacker families (Section~\ref{sec:dataset}). To create a realistic scenario, we generated workloads for the benign processes using their benchmark tools (e.g., \texttt{ab}) and set CPU throttling for the cryptojackers. As shown in Figure~\ref{fig:cpu_benchmark}, detecting cryptojackers is difficult because their CPU usage patterns are similar to (purple line), or even lower (pink line) than, those of benign processes. Additionally, Figure~\ref{fig:pid_obfuscation} demonstrates that the Sysrv-based cryptojacker continuously obfuscates its PID with short and unexpected time intervals, making manual detection and termination ineffective. This example underscores the necessity of developing a new approach to effectively detect and prevent host-based cryptojackers.

\subsection{Existing Solutions}
\label{subsec:existing_work}

To date, various countermeasures have been proposed to combat host-based cryptojackers. In this section, we review existing solutions and evaluate their applicability to detecting and remediating such threats, as well as their suitability for deployment in cloud environments. To guide this analysis, we define five criteria (\textbf{C1}–\textbf{C5}), which are summarized in Table~\ref{table:comparison}.

\noindent\textbf{C1. Supporting Detection.} Most existing detection approaches rely on \emph{dynamic analysis}-based solutions, as static analysis-based methods (e.g., antivirus, blocklist) are ineffective against obfuscation techniques~\cite{tekiner2021sok}. Dynamic analysis-based solutions are broadly categorized into (i) non-machine learning (ML)-based and (ii) ML-based approaches. The former utilizes prominent evidence of cryptomining, such as specific low-level instructions~\cite{lachtar2020rsx} or CPU/RAM usage patterns~\cite{tanana2020wmi}. However, these solutions can be easily bypassed if cryptojackers employ different behavior patterns. In contrast, ML-based approaches aim to enable algorithms to learn cryptojackers' host-level features (e.g., performance counters, syscalls)~\cite{barbhuiya2018rads,ahmad2020dca,mani2020decryptopro,gangwal2020hpc,darabian2020syscall,sanda2023barbhuiya} and network-level features (e.g., packet length)~\cite{tekiner2022ali,carprolu2021noise,gomes2020cryingjackpot}. These solutions are widely used but require collecting fine-grained and extensive features for high accuracy, incurring significant overhead in large-scale cloud environments.

\noindent
\textbf{C2. Supporting Remediation.} While most prior work has centered on detection, relatively little attention has been paid to post-infection remediation strategies. One partial approach~\cite{franco2023honeypot} leverages Suricata, an open-source IDS, to issue alerts based on indicators such as Stratum protocol~\cite{zhang2023under} keywords in DNS requests and the use of ephemeral ports. Although remediation techniques are often not explicitly described, several studies~\cite{ahmad2020dca, gangwal2020hpc, darabian2020syscall, lachtar2020rsx, tanana2020wmi, almurshid2023binary} that focus on process-level detection offer promise, as they identify individual processes that can be terminated. By contrast, other works~\cite{barbhuiya2018rads, mani2020decryptopro, gomes2020cryingjackpot, carprolu2021noise, tekiner2022ali, sanda2023barbhuiya, franco2023honeypot} are limited to detection at the VM level, offering little support for finer-grained remediation.

\begin{table}[t]
\caption{Comparing existing solutions and \ourtool{} for the five criteria. (\protect\fullcirc for fulfillment, \protect\emptycirc for non-compliance)}
\label{table:comparison}
\footnotesize
\centering
\begin{tabular}{cccccc}
\toprule
    \textbf{Work} &
    \textbf{\makecell{C1}} &
    \textbf{\makecell{C2}} & 
    \textbf{\makecell{C3}} &
    \textbf{\makecell{C4}} &
    \textbf{\makecell{C5}} \\ 
    \midrule
Barbhuiya \emph{et al.} (2018)~\cite{barbhuiya2018rads}
    & \fullcirc & \emptycirc & \fullcirc & \emptycirc & \emptycirc \\ \midrule
Ahmad \emph{et al.} (2019)~\cite{ahmad2020dca}
    & \fullcirc & \emptycirc & \fullcirc & \emptycirc & \emptycirc \\ \midrule
Mani \emph{et al.} (2020)~\cite{mani2020decryptopro}
    & \fullcirc & \emptycirc & \emptycirc & \fullcirc & \emptycirc \\ \midrule
Gomes \emph{et al.} (2020)~\cite{gomes2020cryingjackpot}
    & \fullcirc & \emptycirc & \emptycirc & \fullcirc & \emptycirc \\ \midrule
Gangwal \emph{et al.} (2020)~\cite{gangwal2020hpc}
    & \fullcirc & \emptycirc & \fullcirc & \emptycirc & \emptycirc \\ \midrule
Darabian \emph{et al.} (2020)~\cite{darabian2020syscall}
    & \fullcirc & \emptycirc & \emptycirc & \emptycirc & \emptycirc \\ \midrule
Lachtar \emph{et al.} (2020)~\cite{lachtar2020rsx}
    & \fullcirc & \emptycirc & \emptycirc & \fullcirc & \emptycirc \\ \midrule
D. Tanana and G. Tanana (2020)~\cite{tanana2020wmi}
    & \fullcirc & \emptycirc & \emptycirc & \fullcirc & \emptycirc \\ \midrule
Caprolu \emph{et al.} (2021)~\cite{carprolu2021noise}
    & \fullcirc & \emptycirc & \emptycirc & \emptycirc & \emptycirc \\ \midrule
    
Tekiner \emph{et al.} (2022)~\cite{tekiner2022ali}
    & \fullcirc & \emptycirc & \fullcirc & \fullcirc & \emptycirc \\
    \midrule
Sanda \emph{et al.} (2023)~\cite{sanda2023barbhuiya}
    & \fullcirc & \emptycirc & \emptycirc & \emptycirc & \emptycirc \\ \midrule
Almurshid (2023)~\cite{almurshid2023binary}
    & \fullcirc & \emptycirc & \emptycirc & \emptycirc & \emptycirc \\ \midrule
Franco \emph{et al.} (2023)~\cite{franco2023honeypot}
    & \fullcirc & \fullcirc & \emptycirc & \emptycirc & \emptycirc \\ \midrule
\textbf{\ourtool{}} (our work)
    & \fullcirc & \fullcirc & \fullcirc & \fullcirc & \fullcirc \\ \bottomrule
\end{tabular}
\vspace{-0.2in}
\end{table}

\noindent\textbf{C3. Cloud-Ready Deployability.} Existing solutions are often unsuitable for deployment in Linux-based cloud environments. This limitation stems from several factors: some solutions are specifically designed for Windows OS~\cite{mani2020decryptopro, gomes2020cryingjackpot, tanana2020wmi, carprolu2021noise, sanda2023barbhuiya, almurshid2023binary}, while others depend on resource-intensive software~\cite{darabian2020syscall} or hardware infrastructure, such as honeypots~\cite{franco2023honeypot}. Additionally, certain methodologies require custom Linux kernels~\cite{lachtar2020rsx}, making them impractical for the majority of cloud hosts, which typically run off-the-shelf Linux-based systems with limited resources. Furthermore, antivirus programs designed to prevent in-browser cryptojackers~\cite{tekiner2021sok} are unsuitable for cloud environments due to scalability constraints and their proprietary nature (see Appendix~\ref{appen:discussion} for further discussion).

\noindent\textbf{C4. Identifying Obfuscation.} As mentioned, recent host-based cryptojackers use advanced obfuscation techniques to evade conventional detection mechanisms. For instance, \emph{CPU throttling} keeps cryptojackers' CPU usage below a certain threshold, thereby avoiding detection. Several studies~\cite{lachtar2020rsx, tanana2020wmi, mani2020decryptopro, gomes2020cryingjackpot} have proposed countermeasures against this technique. However, determining an appropriate threshold that varies with the execution environment remains a challenge. Additionally, \emph{PID obfuscation}, which involves periodically mutating PIDs, is commonly employed but not addressed in previous solutions.

\begin{figure*}[!t]
    \centering
    \includegraphics[width=.95\textwidth]{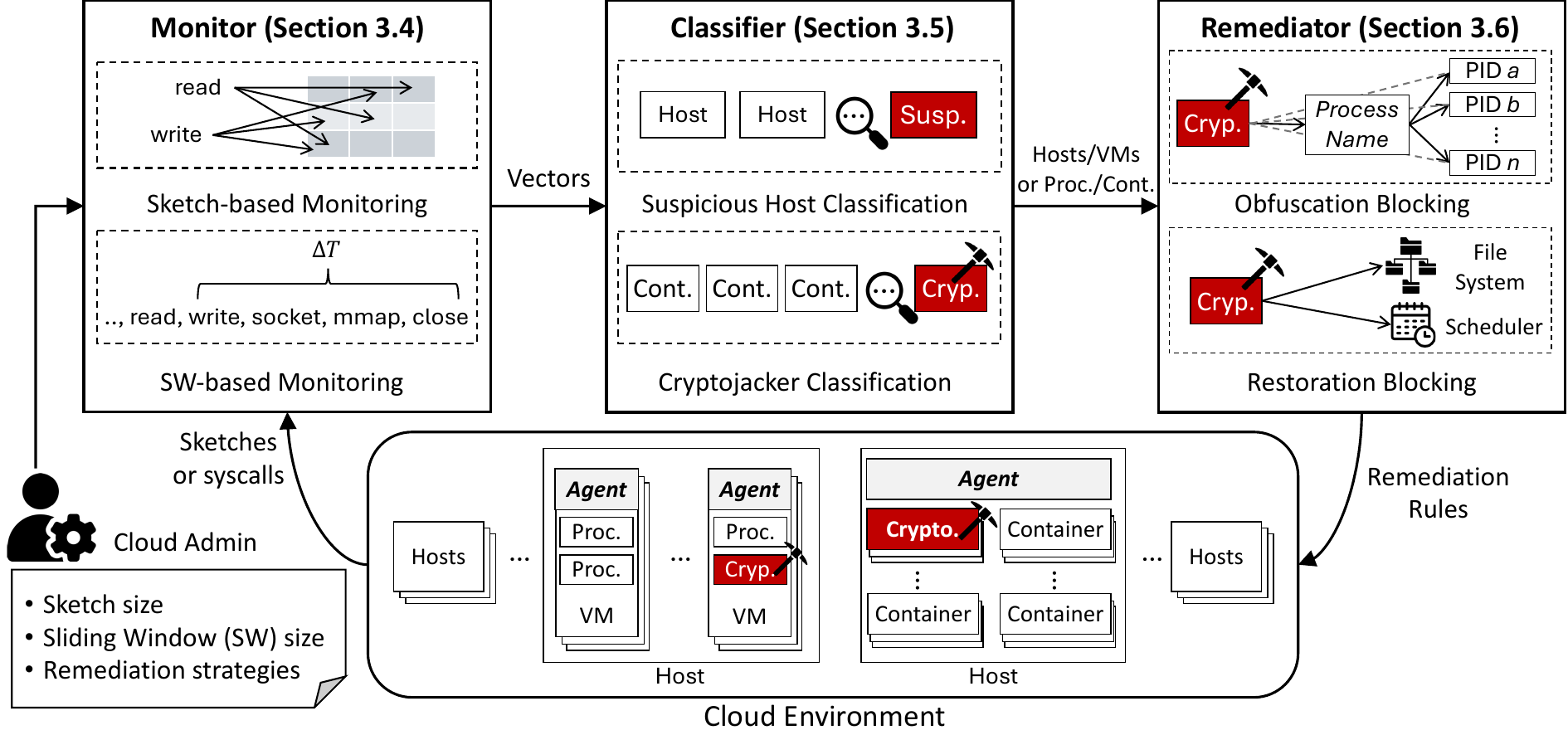}
    \caption{\ourtool{} system architecture and workflow.}
    \label{fig:overview}
    \vspace{-0.15in}
\end{figure*}

\noindent\textbf{C5. Evading Persistence.} Host-based cryptojackers maintain persistence by employing restoration techniques~\cite{vmware_report}. A critical method is tampering with entry points, such as \texttt{cronjob}~\cite{cron1,cron2} or \texttt{rc.local} \cite{rc_local1,rc_local2}, which ensures the cryptojacker is restored even after a system reboot or process termination. However, no prior research has examined these aspects in host-based cryptojackers. \ourtool{} is the first study to address the prevention of persistence.

\section{\ourtool{} Design}

In this section, we present the design of \ourtool{}. We begin by outlining its goals and providing a system overview, followed by a detailed explanation of its modules.

\subsection{Design Goals}

Our aim is to meet all the requirements outlined in Table~\ref{table:comparison} to effectively identify and thwart host-based cryptojackers while developing a solution compatible with cloud environments. To achieve this, \ourtool{} is guided by the following goals (\textbf{G1}-\textbf{G4}):

\noindent\textbf{G1. Integrated Detection/Remediation Hybrid.} Recognizing that segregating detection and remediation solutions increases operational costs, \ourtool{} advocates for an integrated approach. By supporting both detection and remediation as a cohesive solution, \ourtool{} aims to streamline operations and enhance overall effectiveness in countering cryptojackers at the host level.

\noindent\textbf{G2. Cloud-Ready Defense.} Given the prevalence of OS-level virtualization instances (i.e., containers) and the coexistence of VMs, deploying heavyweight host-level countermeasures such as antivirus programs and Intrusion Detection Systems (IDS) across all hosts is costly. Our goal is to ensure that \ourtool{} can be efficiently integrated into current cloud setups, providing robust defense without significant resource consumption.

\noindent\textbf{G3. Lightweight Monitoring.} Detecting host-level cryptojackers requires monitoring host-level behaviors, such as syscalls. However, collecting these raw features from numerous containers or processes in large-scale cloud environments can result in excessive overhead. Therefore, a pivotal design aspect of \ourtool{} is developing a lightweight method for efficiently collecting host-level features, ensuring minimal impact on system performance.

\noindent\textbf{G4. Precise Detection for Stealthy Cryptojackers.} \ourtool{} aims to detect cryptojackers that employ stealth techniques such as CPU throttling and PID obfuscation. These behaviors often result in subtle, short-lived syscall patterns that are difficult to capture using classical ML models based on coarse-grained features. Instead, we employ deep learning models (e.g., CNNs, LSTMs) capable of learning fine-grained temporal dependencies in syscall sequences, enabling robust detection of obfuscated threats.

\subsection{Assumption and Threat Model}

We assume that one or more cryptojackers are operating alongside other processes or containers within a cloud environment, which consists of numerous hosts or VMs. The goal of the cloud administrator is to automatically detect and prevent these cryptojackers from illicitly consuming host resources. These hosts are presumed to have limited resources, similar to standard instances in cloud services (e.g., Amazon EC2 instances), necessitating the design of an efficient system tailored for cloud environments. Additionally, we assume that target hosts are Linux-based systems capable of executing eBPF programs, as eBPF is now a standard feature supported by the Linux kernel. Lastly, our model should not be trained on the cloud host to avoid performance overhead.

It is assumed that a cryptojacker can be downloaded and installed on any cloud host. While this requires attackers to deliver the cryptojacker binary, the specifics of this process are beyond the scope of our paper. Attackers may utilize methods such as dictionary attacks for SSH or Telnet login. We assume the cryptojacker focuses solely on cryptomining and does not execute other types of attacks, such as privilege escalation, as these can be detected by an IDS, hindering the long-term benefits obtained from stealthy cryptomining. Finally, we assume that a cryptojacker can be running as a process or a container; the former aims to target a VM while the latter targets a host where container engines or orchestration platforms operate (e.g., Docker, Kubernetes).

\subsection{System Overview}

To achieve its objectives, \ourtool{} is implemented as a hybrid system that integrates machine learning-based detection with eBPF (extended Berkeley Packet Filter)-driven remediation. Figure~\ref{fig:overview} presents the architecture of \ourtool{}, a flexible and general-purpose framework compatible with modern cloud infrastructures, including both Linux virtual machines (VMs) and containerized environments. \ourtool{} consists of three core modules: \emph{Monitor}, \emph{Classifier}, and \emph{Remediator}. The \emph{Monitor} module supports sketch-based monitoring for hosts and VMs, and sliding window (SW)-based monitoring for containers and processes. It generates feature vectors via lightweight \emph{Agents} deployed at target endpoints for efficient data collection. The \emph{Classifier} performs a two-stage classification using pre-trained machine learning models to identify suspicious VMs, hosts, containers, or processes—enabling effective detection of cryptojackers, including emerging variants. The \emph{Remediator} validates the detection results by examining concrete evidence (e.g., poisoned schedulers or malicious cache directories) to confirm the presence of cryptojackers. It then initiates mitigation by executing tailored countermeasures such as tracking obfuscated PIDs, terminating malicious entities, and removing persistence mechanisms. These capabilities are exposed via APIs, allowing cloud administrators to selectively apply appropriate remediation actions. The following sections describe the design details of each module.

\vspace{-0.1in}
\subsection{Two-Phase Syscall Monitoring} \label{sec:monitoring}

Our focus is on analyzing syscalls from a host-based cryptojacker among various host-level features due to their robustness. The core idea is that syscall patterns tend to remain consistent even when obfuscation techniques are applied. To implement this, we utilize Linux kernel tracepoints to gather syscalls from a host-based cryptojacker. An eBPF program deployed in the kernel space captures and transmits structured data (including PID, UID, process name, and syscall) to user space via a BPF perf buffer. Tracepoints, which are static probes located in the Linux kernel at \texttt{/sys/kernel/tracing/events}, have the advantage of not requiring custom kernel modules~\cite{tracepoint}. Note that our analysis specifically targets syscalls generated from x86-based CPU architectures, excluding other architectures such as ARM.

\noindent\textbf{Tracepoint Subset Monitoring.} It is important to note that tracing all syscalls is impractical due to the significant overhead involved. Additionally, monitoring all syscalls from numerous containers or processes in cloud environments poses a scalability issue. For instance, according to the CNCF 2020 survey~\cite{cncf_2020survey}, 23\% of companies reported running more than 5,000 containers on average. This number has dramatically increased, as evidenced by Datadog's recent analysis of 2.4 billion containers from tens of thousands of their customers~\cite{datadog_report}. Given the steady increase in cloud workloads, the number of generated syscalls will be very high.

To address this, \ourtool{} monitors only the \texttt{enter} syscall tracepoints from the comprehensive set of syscall tracepoints. This approach is based on the observation that when a process invokes a syscall, the \texttt{enter} tracepoint provides crucial context, such as the arguments passed and the state at the moment of invocation. Thus, it is sufficient to trace \texttt{enter} syscalls to learn the behavior of cryptojackers.

\noindent\textbf{Phase-1: Sketch-based Host/VM Monitoring.} Analyzing syscall patterns requires managing data structures for each process or container, which imposes a high overhead. For instance, if \ourtool{} collects syscalls from all processes or containers, the syscall feature vector size is \( F \) for each of the \( S \) syscalls from \( P \) processes (or containers) over a certain period. The estimated space complexity in this case is \( O(F \cdot S \cdot P) \). This high space complexity can be prohibitive in large-scale cloud environments with numerous processes and containers.

\begin{figure}[t]
    \centering
    \includegraphics[width=1\linewidth, trim={0 0.1cm 0 0.1cm}, clip]{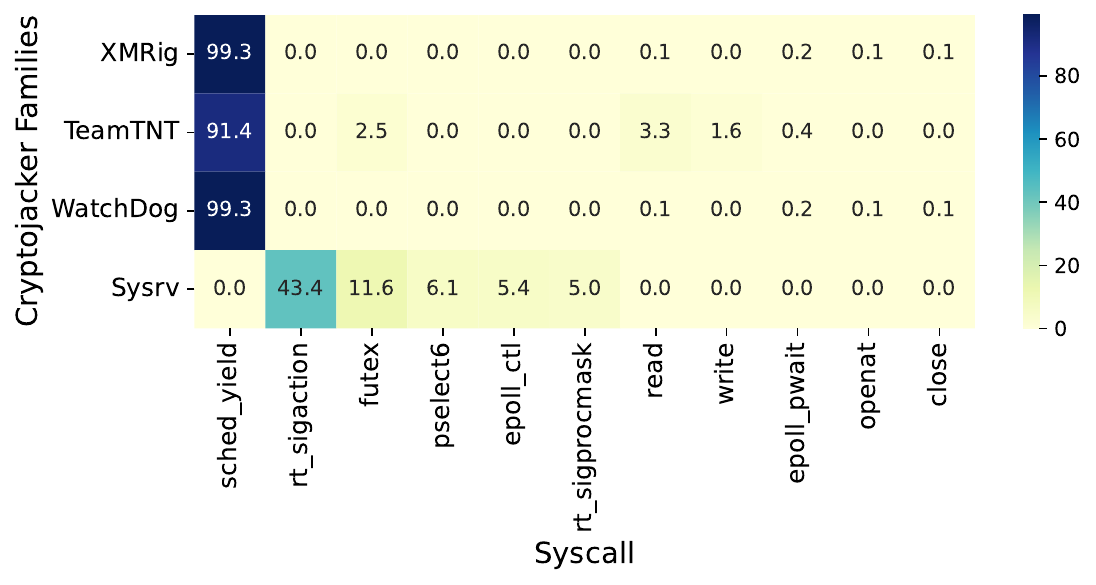}
    \caption{Percentages of the top 5 most frequently invoked syscalls for 4 different cryptojacker families observed in our experiment.}
    \label{fig:notable_syscall}
\vspace{-0.25in}
\end{figure}

To address this challenge, we utilize the Count-Min Sketch (CMS) \cite{cormode2009count}, a probabilistic data structure used for implementing approximate counters with low overhead. Our key idea is to trace syscall distributions for each host or VM instead of all containers or processes. This approach is based on the observation that cryptojackers use specific syscall types, showing unique distributions. As demonstrated in Figure~\ref{fig:notable_syscall}, the top 5 most frequently invoked syscalls from 4 different cryptojacker families reveal distinct patterns. For example, XMRig, TeamTNT, and WatchDog actively use \texttt{sched\_yield}, while Sysrv predominantly uses \texttt{rt\_sigaction}. Although these syscalls can also be used by benign processes, our hypothesis is that cryptojackers are more likely to invoke these syscalls.

To this end, \ourtool{} Agent collects CMSs that represent syscall distributions. Formally, each CMS instance per host or VM is defined as a matrix:
\[
\text{CMS}_h(t) = \begin{bmatrix}
c_{1,1} & c_{1,2} & \ldots & c_{1,w} \\
c_{2,1} & c_{2,2} & \ldots & c_{2,w} \\
\vdots & \vdots & \ddots & \vdots \\
c_{d,1} & c_{d,2} & \ldots & c_{d,w}
\end{bmatrix},
\]
where \( h \) is a host or a VM, \( t \) is the timestamp, and \( c_{i,j} \) represents the count of items hashed to the \( j \)-th bucket by the \( i \)-th hash function. The estimated space complexity is \( O(H \cdot w \cdot d) \), where \( H \) is the number of hosts or VMs, and \( w \) and \( d \) are the width and depth of the sketch, respectively.

Typically, the width \( w \) is determined by the desired error rate (\( \epsilon \)) and is given by \( w = \lceil e / \epsilon \rceil \), while the depth \( d \) is determined by the confidence level (\( \delta \)) and is given by \( d = \lceil \ln (1 / \delta) \rceil \). For example, setting the desired error rate \( \epsilon = 0.01 \) and confidence level \( \delta = 0.1 \) results in \( w = 272 \) and \( d = 3 \). Thus, in large-scale environments where \( P \) is significantly larger than \( H \), and considering that the values of \( w \) and \( d \) are relatively small and constant compared to \( S \) and \( P \), the space complexity without sketches is much higher than with sketches. By using this sketch-based monitoring, we can significantly reduce space complexity, making it more feasible for large-scale cloud environments.

\noindent\textbf{Phase-2: Sliding Window-based Container/Process Monitoring.} When a certain host or VM is classified as suspicious (Section~\ref{sec:detection}), we need to investigate syscall patterns for all processes and containers on that host in detail to pinpoint cryptojackers. Each host agent maintains syscall traces that are used for creating sketches, but analyzing all these traces over the entire timeline requires significant overhead. To address this, \ourtool{} Agents keep track of the recent \( \Delta T \) time window of syscalls using a sliding window. The idea behind this approach is that cryptojackers are likely to operate consistently over a certain period. Thus, when \ourtool{} detects a suspicious host, analyzing the recent syscalls within the time window \( \Delta T \) is sufficient to pinpoint a cryptojacker process or container within the host if the window size \( \Delta T \) is carefully chosen. This way, we can ensure low inference time and space complexity.

\subsection{Two-Phase Cryptojacking Detection} \label{sec:detection}

This section explains how \ourtool{} employs deep learning to detect host-based cryptojackers. In cloud environments, identifying cryptojacker processes or containers is challenging due to the large number of hosts and VMs, each running numerous benign processes. This complexity often requires cloud administrators to manually scan for cryptojackers across an extensive search space, which is time-consuming. To address this issue, we decompose the detection process into two phases: (i) classifying suspicious hosts or VMs, and (ii) classifying cryptojacker containers or processes.

\noindent\textbf{Phase-1: Classifying Suspicious Hosts or VMs.} The goal of this phase is to identify suspicious hosts or VMs that may contain cryptojackers. As discussed in Section~\ref{sec:monitoring}, our hypothesis is that suspicious hosts or VMs are likely to have different syscall distributions compared to benign hosts or VMs. By identifying these suspicious hosts, we can narrow down the large search space of cloud environments. As mentioned, \ourtool{} collects Count-Min Sketches (CMS) from hosts or VMs. 

To apply CMS features to deep learning models, we transform each CMS into a feature vector. Intuitively, for each host \( h \) and timestamp \( t \), we estimate the frequency of \( k \) distinct syscalls \( s_i \in \{s_1, \ldots, s_k\} \) using the CMS sketch. Each frequency is obtained by taking the minimum count across multiple hash functions. The resulting \( k \) estimated frequencies form a fixed-length feature vector \( \mathbf{v}_h(t) \in \mathbb{R}^k \), where each dimension corresponds to a specific syscall. This transformation compresses raw syscall activity into a compact numerical representation suitable for deep learning. For example, if we track \( k = 100 \) syscall types, we obtain a 100-dimensional feature vector where each element reflects the estimated frequency of a corresponding syscall at timestamp \( t \).

For a given syscall \( s \in S \), the estimated frequency \( \hat{f}(s) \) is computed as:
\[
\hat{f}(s, t) = \min_{1 \leq i \leq d} \text{CMS}_h(t)[i][h_i(s)],
\]
where \( h_i(s) \) is the bucket index for the \( i \)-th hash function. To construct the feature vector \( \mathbf{v}_h(t) \) for \( k \) distinct syscalls \( \{s_1, s_2, \ldots, s_k\} \), we compute:
\[
\mathbf{v}_h(t) = [\hat{f}(s_1, t), \hat{f}(s_2, t), \ldots, \hat{f}(s_k, t)],
\]
which represents the distribution of syscalls on host \( h \) during the time interval \( t \). We extend this feature vector across multiple time intervals \( t_1, t_2, \ldots, t_n \). For \( k \) distinct syscalls and \( n \) time intervals, we construct a time-series feature matrix \( \mathbf{V}_h \) as follows:
\[
\mathbf{V}_h = \begin{bmatrix}
\mathbf{v}_h(t_1) \\
\mathbf{v}_h(t_2) \\
\vdots \\
\mathbf{v}_h(t_n)
\end{bmatrix} = \begin{bmatrix}
\hat{f}(s_1, t_1) & \hat{f}(s_2, t_1) & \ldots & \hat{f}(s_k, t_1) \\
\hat{f}(s_1, t_2) & \hat{f}(s_2, t_2) & \ldots & \hat{f}(s_k, t_2) \\
\vdots & \vdots & \ddots & \vdots \\
\hat{f}(s_1, t_n) & \hat{f}(s_2, t_n) & \ldots & \hat{f}(s_k, t_n)
\end{bmatrix}.
\]

We conduct a binary classification through a deep learning model \( \mathcal{M}_{phase1} \) that outputs a probability score \( \hat{y} \) for the presence of cryptojackers. The model is trained using a labeled dataset \( \{(\mathbf{V}_h, y_h)\} \), where \( y_h \in \{0, 1\} \) indicates the absence or presence of cryptojackers on a host or VM \( h \). The binary cross-entropy (BCE) loss function is used to train the model, defined as follows:
\[
\mathcal{L}_{BCE} = -\frac{1}{N} \sum_{h=1}^{N} \left[ y_h \log(\hat{y}_h) + (1 - y_h) \log(1 - \hat{y}_h) \right],
\]
where \( N \) is the number of hosts, \( \hat{y}_h \) is the predicted probability for a host \( h \), and \( y_h \) is the ground truth label. By minimizing this loss function during training, the model learns to distinguish between hosts or VMs with and without malware based on their syscall distributions over time.

\noindent\textbf{Phase-2: Classifying Cryptojacker Containers or Processes.} After identifying a suspicious host or VM, the next goal is to detect the specific cryptojacker and determine its family among the various containers or processes. Different cryptojacker families exhibit unique syscall patterns, which often require tailored remediation strategies (Section~\ref{sec:prevention}). To ensure efficiency, we focus on analyzing the most recent \( \Delta T \) time window of syscalls extracted using the sliding window mechanism (Section~\ref{sec:monitoring}). Let \( \mathbf{s}_p \) represent the syscall sequence observed during the time window \( \Delta T \) at time \( t \), formalized as follows:
\[
\mathbf{s}_p(t, \Delta T) = [s_{t-\Delta T+1}, s_{t-\Delta T+2}, \ldots, s_t],
\]
where \( p \) is the process or container and \( s_p \) is the observed syscall at a time \( t \). Then, we define a syscall embedding \( \mathbf{e}_{s_t} \) as:
\[
\mathbf{e}_{s_t} = E[s_t],
\]
where \( E \in \mathbb{R}^{k \times d} \) is a lookup embedding matrix for embedding  \( k \) distinct syscall types into \( d \)-dimensional vectors. As the input feature of the process or container \( p \), the sequence of syscall embeddings is formalized as follows:
\[
\mathbf{v}_p(t) = [\mathbf{e}_{s_{t-\Delta T+1}}, \mathbf{e}_{s_{t-\Delta T+2}}, \ldots, \mathbf{e}_{s_t}].
\]

This syscall sequence feature is fed into a deep learning model \( \mathcal{M}_{phase2} \) to capture temporal syscall patterns. We aim to classify whether a process or container \( p \) belongs to a specific cryptojacker family. To do so, we conduct a multi-class classification to identify the cryptojacker family associated with the process or container \( p \). Through the model \( \mathcal{M}_{phase2} \), a prediction probability \( \hat{y}_{p,c} \) for each cryptojacker family \( c \in \{1, 2, \ldots, C\} \), where \( C \) is the number of cryptojacker families, is derived from the syscall representation vector. Based on the ground truth label \( y_{p} \), the embedding vectors and deep learning model are trained in an end-to-end approach by optimizing the cross-entropy (CE) loss:
\[
\mathcal{L}_{CE} = -\frac{1}{N} \sum_{p=1}^{N} \sum_{c=1}^{C} y_{p} \log(\hat{y}_{p,c}),
\]
where \( N \) is the number of processes or containers, \( y_{p} \) is the ground truth label, and \( \hat{y}_{p,c} \) is the predicted probability for class \( c \) (i.e., cryptojacker family). By applying these deep learning-based classification methods, we can accurately classify each process or container as a specific cryptojacker family based on their syscall sequences.

\begin{figure}[t]
    \centering
    \includegraphics[width=\linewidth]{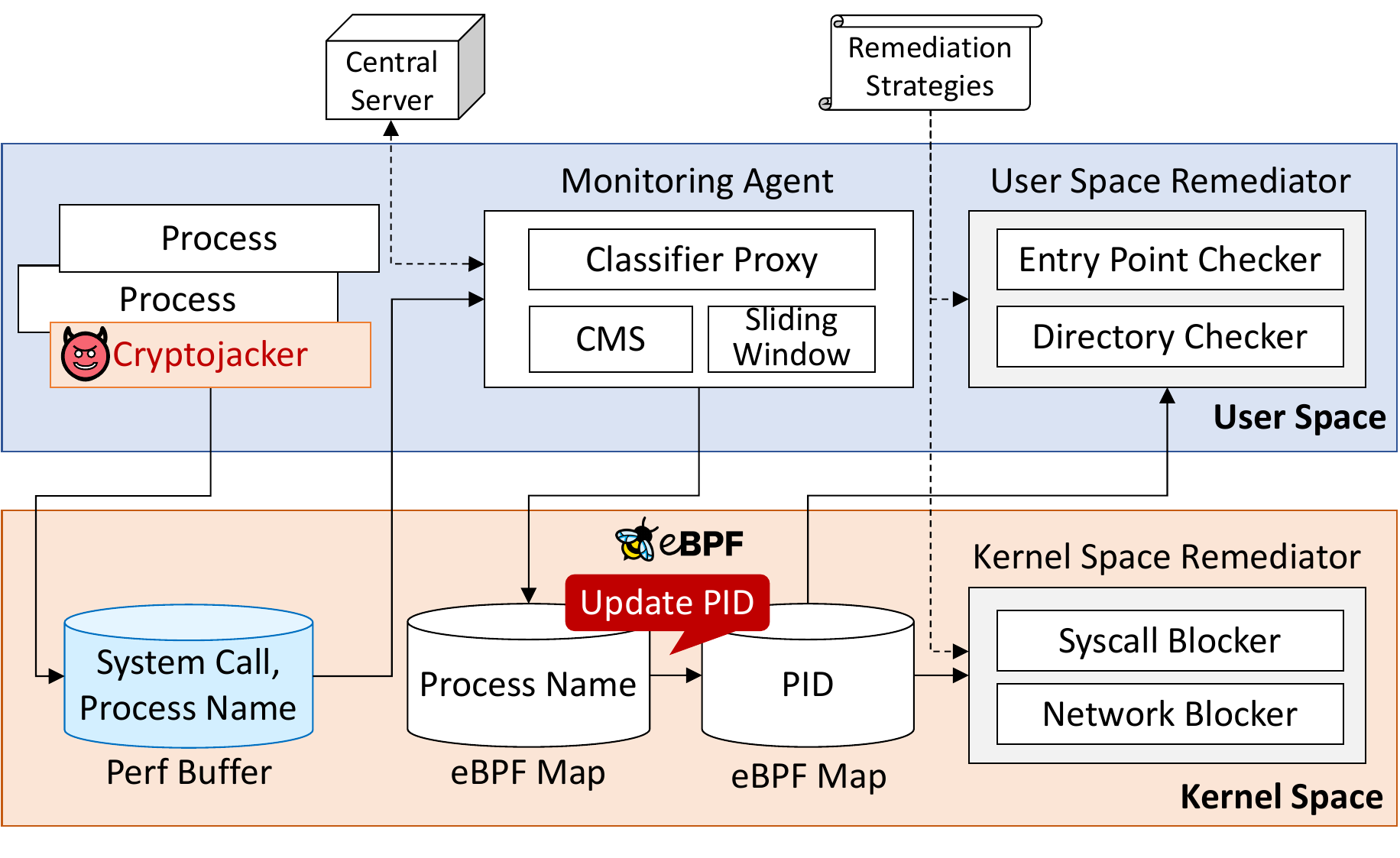}    \caption{Workflow of modules in CryptoGuard Agent.}
    \label{fig:overview_2}
    \vspace{-0.1in}
\end{figure}

\subsection{Cryptojacking Remediation} \label{sec:prevention}

We now describe the methodologies employed by \ourtool{} to remediate hosts compromised by identified cryptojackers.

\noindent
\textbf{Workflow.} Remediating cryptojacking requires seamless coordination among multiple components of the \ourtool{} Agent deployed on each host or VM. Figure~\ref{fig:overview_2} illustrates the internal architecture of the \ourtool{} Agent. The Monitoring Agent communicates with \ourtool{}'s central server by transmitting either Count-Min Sketch (CMS) data or sliding window syscall traces. When the Classifier at the central server identifies a cryptojacker process, it relays the corresponding process name to the Monitoring Agent via the Classifier Proxy. The Monitoring Agent then updates an eBPF map with this information. To reduce the risk of false positives, the User Space Remediator performs an additional validation step by checking whether the flagged process is registered in entrypoints or cache directories—locations commonly abused by host-based cryptojackers for persistence~\cite{cron1,cron2,rc_local1,rc_local2}. In parallel, the process's PID is recorded in a separate eBPF map to maintain accurate \emph{name--PID pairs}, which are critical for preventing the cryptojacker from re-establishing itself on the system. Based on this information, the User Space Remediator or Kernel Space Remediator modules execute remediation strategies configured by the cloud administrator. \ourtool{} supports a range of remediation actions, as detailed below:

\noindent\textbf{Blocking Network Connections.}  
Cryptojackers depend on persistent network connections with mining pools to receive mining jobs and submit discovered hashes~\cite{tekiner2021sok}. While blocking connections to publicly known mining pools (e.g., Nanopool~\cite{nanopool}) or widely used protocols (e.g., Stratum) may seem straightforward, modern cryptojackers have adapted by utilizing private mining pools, custom ports, and SSL/TLS encryption to evade detection~\cite{zhang2023under}. As a result, static rules targeting known signatures are often ineffective. To counter this, it is crucial to dynamically update remediation rules to block cryptojackers from establishing similar connections. \ourtool{} achieves this by maintaining eBPF maps in kernel space to manage identified process names. It employs Traffic Control (TC)~\cite{tc-bpf}, which operates after the execution of XDP for network packet filtering~\cite{Network_eBPF}. By leveraging TC, \ourtool{} filters outgoing (TX side) packets, dropping TCP packets associated with flagged processes or containers at the kernel level, effectively preventing cryptojackers from maintaining connections to mining pools and profiting further.

\noindent\textbf{Blocking Popular Syscalls.} Another effective way to thwart cryptojackers is by blocking specific syscalls to limit their functionality. This approach offers advantages over simply killing or removing processes, as it allows cloud administrators to investigate suspicious processes or containers, observe their behavior, and perform root cause analysis while cryptojackers are quarantined. However, blocking all syscalls of suspicious processes or containers is inefficient. Instead, \ourtool{} focuses on popular syscalls commonly invoked by cryptojackers. For example, we have observed that specific syscalls (e.g., \texttt{sched\_yield}, \texttt{rt\_sigaction}) are frequently used by different families of cryptojackers (Figure~\ref{fig:notable_syscall}). XMRig-based cryptojackers use the \texttt{sched\_yield} syscall to cause the calling thread to relinquish the CPU, allowing other processes to use the CPU and thereby keeping the cryptojacker's CPU usage low. Sysrv-based cryptojackers use the \texttt{rt\_sigaction} syscall to change or set the signal handler associated with a specific action. Both syscalls are employed for the obfuscation and restoration of cryptojackers. When these syscalls are blocked, a cryptojacker may not function correctly, potentially leading to its termination or making it evident through detectable behaviors such as high CPU usage.

\noindent
\textbf{Killing Cryptojackers.} Prior approaches have mitigated cryptojackers by suspending~\cite{bian2020minethrottle} or terminating~\cite{kill} their processes. However, as noted in Section~\ref{sec:motivation}, modern cryptojackers often employ PID obfuscation, rapidly changing their process IDs to evade such countermeasures. To overcome this, \ourtool{} extracts the \textit{process name} of identified cryptojackers and stores it in an eBPF map. It then periodically issues SIGKILL signals to all processes or containers matching these names, ensuring that all instances are terminated. This strategy leverages the insight that, unlike PIDs, process names tend to remain stable. To minimize disruption from false positives, \ourtool{} verifies supporting evidence—such as poisoned entrypoints or malicious cache directories—before initiating termination.

\begin{figure}[!t]
    \centering
    \subfloat[Examples of a crontab file poisoned by cryptojackers. The first five fields indicate the execution frequency, and the last field specifies the command for executing a cryptojacker binary. These entries are removed after being detected from syscall arguments.]{
        \includegraphics[width=1\linewidth]{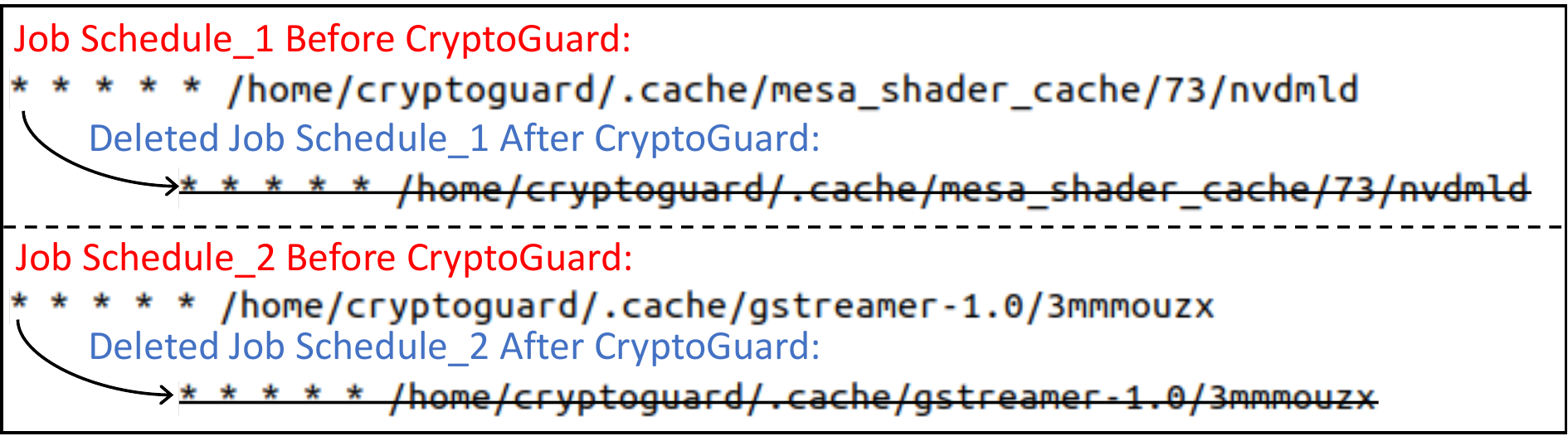}
        \label{fig:cron_example}
    }

    \subfloat[Analysis of syscall arguments involving a cache directory where a cryptojacker binary is injected.]{
        \includegraphics[width=1\linewidth]{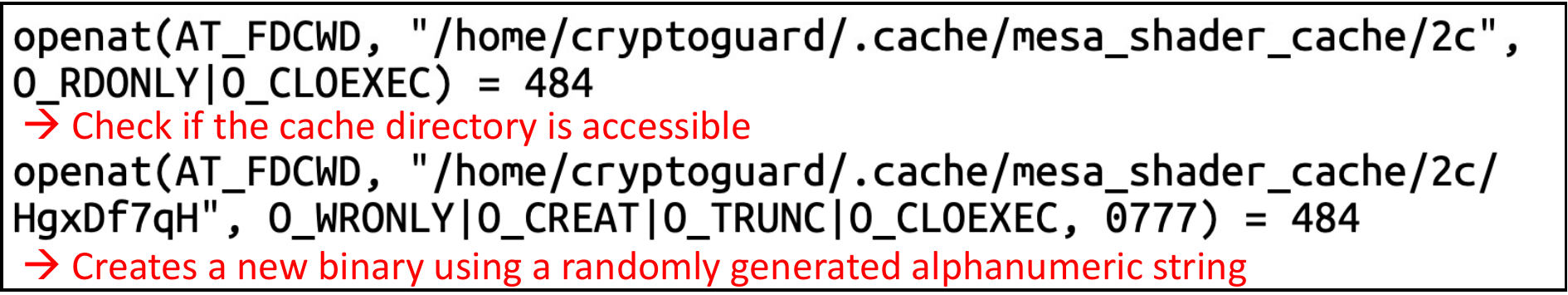}
        \label{fig:cache_example}
    }
    \caption{Examples of thwarting cryptojackers' persistence.}
\vspace{-0.15in}
\end{figure}

\noindent\textbf{Thwarting Persistence.}
If cryptojackers are installed in VMs or bare-metal hosts, they tend to exploit various entry points to remain persistent. For example, in our tests involving 123 cryptojacker samples (Section~\ref{sec:dataset}), we noticed that 40 of these samples ensured the periodic restoration of the cryptojacker by utilizing \texttt{cron}, a Linux job scheduler. Figure~\ref{fig:cron_example} illustrates a crontab file poisoned by a cryptojacker when it is installed on the host. In addition, reports~\cite{rc_local1,rc_local2} reveal that some cryptojackers compromise Linux \texttt{rc.local} to execute themselves at start-up time. Thus, simply terminating cryptojacker processes is ineffective as long as such entries remain. To address this, \ourtool{} focuses on the deployment mechanism of cryptojackers and their syscall arguments. Figure~\ref{fig:cache_example} shows that a cryptojacker calls the \texttt{openat} syscall to check whether the cache directory is accessible and, if so, injects a new binary using a randomly generated alphanumeric string. Here, the syscall arguments contain the target directory and binary path. By referencing this information, \ourtool{} can remove not only the binary but also the entrypoints from the scheduler (i.e., crontab), preventing the cryptojacker from restoring its activities.

\noindent\textbf{Supporting High-level APIs.}
Remediation strategies can vary depending on the specific needs and preferences of system administrators. To accommodate these diverse requirements, \ourtool{} offers a set of high-level APIs that enable the implementation of customized remediation workflows. Further details are provided in Appendix~\ref{appen:algorithm}.

\section{Evaluation}

In this section, we evaluate \ourtool{} to demonstrate its effectiveness and efficiency.

\subsection{Implementation}
\label{subsec:implementation}

We implemented \ourtool{} with 5,000 lines of C code, leveraging several advanced tools and techniques. To develop \ourtool{}'s Monitor and Agent, we utilized bpftrace~\cite{bpftrace_tool}, which we found particularly well-suited compared to alternatives like SystemTap, LTTng, Perf, and DTrace. The suitability of bpftrace stemmed from its: (i) faster program creation and execution enabled by eBPF, (ii) support for eBPF maps for key/value data structures, and (iii) program safety ensured by the eBPF verifier~\cite{bpftrace}. For our Classifiers, we used TensorFlow and designed two deep learning models—(i) Convolutional Neural Network (CNN) and (ii) Long Short-Term Memory (LSTM)—due to their effectiveness in learning time-series patterns~\cite{seo2022heimdallr,cao2019fingerprinting}. 
Figure~\ref{fig:model_architecture} in Appendix~\ref{appen:model_arch} shows the deep learning model architectures for the binary classification task in Phase-2. For Phase-1, the output dimension of the last fully connected layer in each model for the binary classification task differs to that in the presented example.
The Remediator was implemented using eBPF programs built with libbpf, and \ourtool{} was integrated with Kubernetes for deployment in a containerized environment. Specifically, we employed DaemonSet to automatically deploy the \ourtool{} Agent to each Kubernetes node as a pod, enabling syscall monitoring across all nodes. In Phase-1 mode, the \ourtool{} Agent collects a sketch for each node, while in Phase-2, it picks a container and collects its syscalls using a sliding window.

\subsection{Experimental Environment}
\label{sec:environment}

The experiments were performed on a machine equipped with an Intel Xeon Gold 5220R CPU running at 2.20GHz and 200GB of RAM. This machine utilized the KVM (Kernel-based Virtual Machine) hypervisor, through which we created 50 VMs using libvirt to establish the testbed for data collection. Additionally, our deep learning models were trained on a separate machine featuring two NVIDIA Tesla V100 GPUs, and 300GB of RAM.

\subsection{Dataset Collection}
\label{sec:dataset}

\noindent
\textbf{Benign Samples.}
To construct our dataset, we emulated a realistic deployment environment by running commonly used benign services, including Apache HTTP Server, Apache Tomcat, Nginx, Redis, and MySQL. To mimic real-world usage patterns, we applied substantial workloads—up to 100 million HTTP requests using up to 10 threads—through benchmarking tools such as \texttt{ab} and \texttt{sysbench}. This setup enabled us to capture a broad range of benign system call sequences representative of typical server behavior.

\noindent\textbf{Cryptojacker Samples.}  
We utilized real cryptojacker samples published in the VMware Threat Report 2022~\cite{vmware_report} by the VMware NSX Threat Intelligence Team. These samples were selected for their reliability, having been curated by the VMware Team from VirusTotal between June 2021 and November 2021. The files are in the ELF 64-bit LSB executable format for x86-64 architecture. The dataset includes a variety of cryptojacker families, such as XMRig~\cite{xmrig}, Sysrv~\cite{sysrv}, TeamTNT~\cite{teamtnt}, Mexalz~\cite{mexalz}, Omelette~\cite{omelette}, WatchDog~\cite{watchdog}, and Kinsing~\cite{kinsing}. Of the 139 cryptojacker samples accessed, we selected 123 ELF files for our research, excluding those that caused segmentation faults or failed to operate within our testbed. A detailed distribution of cryptojacker samples across families is provided in Table~\ref{table:cryptojacker_family}. Notably, we categorized XMRig, TeamTNT, and WatchDog under the \emph{XMRig} group due to their similar behavioral patterns, while the remaining samples are grouped as \emph{Sysrv} families.

\begin{table}[t]
\caption{Breakdown of cryptojacker samples in~\cite{vmware_samples}.}
\vspace{-0.1in}
\label{table:cryptojacker_family}
\centering
\footnotesize
\begin{tabular}{cccc}
\toprule
    \textbf{Family} & \textbf{\makecell{Total Samples}} & \textbf{\makecell{Testbed Executable}} & \textbf{Class}\\ 
    \midrule
    Sysrv & 100 & 100 & Sysrv \\ 
    \midrule
    XMRig & 17 & 15 & \multirow{3}{*}{XMRig}  \\ 
    TeamTNT & 7 & 5 & \\ 
    WatchDog & 3 & 3 &\\ 
    \midrule
    Mexalz & 5 & 0 & \multirow{3}{*}{N/A} \\
    Omelette & 4 & 0 & \\ 
    Kinsing & 3 & 0 & \\
    \midrule
    Total & 139 & 123 & \\
    \bottomrule
\end{tabular}
\vspace{-0.25in}
\end{table}

\noindent\textbf{Collection Methodology.} 
We collected syscalls invoked by each process over a 300-second period after its execution. The data for Count-Min Sketches (CMS) and syscall sequences for each process was saved in individual CSV files. Processes were differentiated based on their process names, as defined by the length specified in \texttt{TASK\_COMM\_LEN}, a constant in the Linux kernel that sets the maximum allowed limit on the length of a process name. In our case, it was set to 16.

\noindent\textbf{Phase-1 Dataset.} We collected a Phase-1 dataset comprising 4,020 entries collected from each VM in our testbed. Using benchmark tools and processes, we collected 2,010 benign CMS datasets. Under the same conditions, we collected an additional 2,010 malware CMS datasets from an environment where a cryptojacker was actively running. To effectively evaluate the model’s generalization ability, we split both the benign and malware CMS datasets based on their timeframes: 0–168 seconds for training, 169–210 seconds for validation, and 211–300 seconds for testing. This split allows us to verify whether or not our deep learning models can detect unseen CMS patterns.

\noindent\textbf{Phase-2 Dataset.} For Phase-2 dataset, we used sliding windows whose size is $\Delta T = 30$, 60, and 90 seconds. Then, we collected 2,131, 1,219, and 924 samples of benign datasets, respectively. For the Xmrig cryptojacker class, we collected 500, 350, and 310 datasets, and for the Sysrv class, we collected 1,059, 469, and 352 datasets, respectively. For each split from the Phase-1, we use 2,250 samples for training, 562 samples for validation, and 1,206 samples for testing.

\begin{table*}[t]
\caption{Phase-1 classification results for detecting suspicious hosts.}
\vspace{-0.1in}
\label{table:phase1_results}
\scriptsize
\centering
\begin{tabular}{cc ccc ccc ccc}
\toprule
    \multirow{3}{*}{\makecell{\textbf{CMS Size}\\(width $\times$ depth)}} & \multirow{3}{*}{\textbf{Class}} & \multicolumn{3}{c}{\textbf{Random Forest (RF)}} & \multicolumn{3}{c}{\textbf{LSTM}} & \multicolumn{3}{c}{\textbf{CNN}} \\ \cmidrule(lr){3-5} \cmidrule(lr){6-8} \cmidrule(lr){9-11}
    &  & Precision & Recall & F1-score & Precision & Recall & F1-score & Precision & Recall & F1-score \\ 
    \midrule

    \multirow{3}{*}{\makecell{$272 \times 3$\\$(\epsilon=0.01, \delta=0.1)$}}
    & Benign & 0.9736 & 0.9439 & 0.9585 & 0.9462 & 0.9888 & 0.9670 & 0.9736 & 0.9439 & 0.9585\\
    & Malware & 0.9418 & 0.9725 & 0.9569 & 0.9874 & 0.9400 & 0.9631 & 0.9418 & 0.9725 & 0.9569\\ \cmidrule(lr){3-5} \cmidrule(lr){6-8} \cmidrule(lr){9-11}

    & Average & 0.9577 & 0.9582 & 0.9577 & \textbf{0.9668} & \textbf{0.9644} & \textbf{0.9651} & 0.9557 & 0.9582 & 0.9577 \\\midrule

    \multirow{3}{*}{\makecell{$55 \times 3$\\$(\epsilon=0.01, \delta=0.01)$}}
    & Benign & 0.9583 & 0.9567 & 0.9575 & 0.9682 & 0.9621 & 0.9651 & 0.9543 & 0.9528 & 0.9536\\
    & Malware & 0.9537 & 0.9553 & 0.9545 & 0.9584 & 0.9651 & 0.9617 & 0.9475 & 0.9491 & 0.9483\\ \cmidrule(lr){3-5} \cmidrule(lr){6-8} \cmidrule(lr){9-11}

    & Average & 0.9560 & 0.9560 & 0.9560 & \textbf{0.9633} & \textbf{0.9636} & \textbf{0.9634} & 0.9509 & 0.9510 & 0.9510 \\\midrule

    \multirow{3}{*}{\makecell{$55 \times 5$\\$(\epsilon=0.05, \delta=0.01)$}}
    & Benign & 0.9659 & 0.9535 & 0.9596 & 0.9567 & 0.9763 & 0.9664 & 0.9477 & 0.9888 & 0.9678\\
    & Malware & 0.9312 & 0.9552 & 0.9431 & 0.9732 & 0.9511 & 0.9620 & 0.9874 & 0.9417 & 0.9640\\ \cmidrule(lr){3-5} \cmidrule(lr){6-8} \cmidrule(lr){9-11}

    & Average & 0.9486 & 0.9544 & 0.9514 & 0.9650 & 0.9637 & 0.9642 & \textbf{0.9676} & \textbf{0.9653} & \textbf{0.9659} \\\midrule

    \multirow{2}{*}{Average} & Benign & 0.9659 & 0.9514 & 0.9585 & 0.9570 & 0.9757 & 0.9662 & 0.9585 & 0.9618 & 0.9600\\ 
    & Malware & 0.9422 & 0.9610 & 0.9515 & 0.9730 & 0.9521 & 0.9623 & 0.9589 & 0.9544 & 0.9564\\

    \bottomrule
\end{tabular}
\vspace{-0.1in}
\end{table*}

\begin{table*}[t]
\caption{Phase-2 classification results for detecting cryptojacker processes.}
\vspace{-0.1in}
\label{table:phase2_results}
\scriptsize
\centering

\begin{tabular}{cc ccc ccc ccc}
\toprule

    \multirow{3}{*}{\textbf{Sliding Window Size}} & \multirow{3}{*}{\textbf{Class}} & \multicolumn{3}{c}{\textbf{Random Forest (RF)}} & \multicolumn{3}{c}{\textbf{LSTM}} & \multicolumn{3}{c}{\textbf{CNN}} \\ \cmidrule(lr){3-5} \cmidrule(lr){6-8} \cmidrule(lr){9-11}
    &  & Precision & Recall & F1-score & Precision & Recall & F1-score & Precision & Recall & F1-score \\ 
    \midrule

    \multirow{4}{*}{\makecell{$\Delta T=30$}} 
    & XMRig & 0.8239 & 0.9357 & 0.8763 & 0.7500 & 0.9214 & 0.8269 & 0.9776 & 0.9357 & 0.9562\\
    & Sysrv & 0.9259 & 0.8117 & 0.8651 & 0.9150 & 0.9091 & 0.9121 & 0.9870 & 0.9838 & 0.9854\\ 
    & Benign & 0.9325 & 0.9592 & 0.9456 & 0.9810 & 0.9366 & 0.9583 & 0.9836 & 0.9940 & 0.9887\\ \cmidrule(lr){3-5} \cmidrule(lr){6-8} \cmidrule(lr){9-11}
    & Average & 0.8941 & 0.9022 & 0.8957 & 0.8820 & 0.9224 & 0.8991 & \textbf{0.9827} & \textbf{0.9712} & \textbf{0.9768} \\ \midrule

    \multirow{4}{*}{\makecell{$\Delta T=60$}}
    & XMRig & 0.7368 & 0.8660 & 0.7962 & 0.6911 & 0.8763 & 0.7727 & 0.9362 & 0.9072 & 0.9215\\
    & Sysrv & 0.9286 & 0.7548 & 0.8327 & 0.9085 & 0.8968 & 0.9026 & 0.9934 & 0.9742 & 0.9837\\ 
    & Benign & 0.9154 & 0.9444 & 0.9297 & 0.9576 & 0.8968 & 0.9262 & 0.9766 & 0.9921 & 0.9843\\ \cmidrule(lr){3-5} \cmidrule(lr){6-8} \cmidrule(lr){9-11}
    & Average & 0.8603 & 0.8551 & 0.8529 & 0.8524 & 0.8900 & 0.8672 & \textbf{0.9687} & \textbf{0.9578} & \textbf{0.9631} \\ \midrule

    \multirow{4}{*}{\makecell{$\Delta T=90$}} 
    & XMRig & 0.8710 & 0.7714 & 0.8182 & 0.7025 & 0.8763 & 0.7798 & 0.9688 & 0.9208 & 0.9442\\
    & Sysrv & 0.8523 & 0.7353 & 0.7895 & 0.8875 & 0.9161 & 0.9016 & 0.9519 & 0.9802 & 0.9659\\ 
    & Benign & 0.8691 & 0.9522 & 0.9088 & 0.9656 & 0.8915 & 0.9271 & 0.9677 & 0.9747 & 0.9712\\ \cmidrule(lr){3-5} \cmidrule(lr){6-8} \cmidrule(lr){9-11}
    & Average & 0.8641 & 0.8196 & 0.8388 & 0.8519 & 0.8946 & 0.8695 & \textbf{0.9628} & \textbf{0.9586} & \textbf{0.9604} \\ \midrule
    
    \multirow{3}{*}{Average} & XMRig & 0.8106 & 0.8577 & 0.8302 & 0.7145 & 0.8913 & 0.7931 & 0.9608 & 0.9212 & 0.9208 \\
    & Sysrv & 0.9023 & 0.7673 & 0.8291 & 0.9037 & 0.9073 & 0.9054 & 0.9774 & 0.9794 & 0.9783 \\
    & Benign & 0.9057 & 0.9519 & 0.9280 & 0.9681 & 0.9083 & 0.9372 & 0.9759 & 0.9869 & 0.9814 \\
    
    \bottomrule
\end{tabular}
\vspace{-0.15in}
\end{table*}

\subsection{Classifier Performance}
\label{subsec:classifer_performance}

We conducted experiments to evaluate the performance of our deep learning classifiers in detecting cryptojackers. As a baseline, we employed Random Forest (RF) as a traditional machine learning model to compare with our LTSM and CNN models. We assumed a practical scenario where a cryptojacker process is running with other benign processes at the same host (or VM) concurrently.

\noindent\textbf{Effectiveness of Phase-1 Classification.}
We evaluated the effectiveness of our Phase-1 classifiers, which perform binary classification to distinguish suspicious hosts from benign ones (Section~\ref{sec:detection}). Each VM was labeled as either (i) \textit{Benign} or (ii) \textit{Malware}, and classification was conducted using three different CMS sizes: $272 \times 3$, $55 \times 3$, and $55 \times 5$, corresponding to the configurations $(\epsilon=0.01, \delta=0.1)$, $(\epsilon=0.01, \delta=0.01)$, and $(\epsilon=0.05, \delta=0.01)$, respectively. As shown in Table~\ref{table:phase1_results}, our LSTM and CNN classifiers achieved average F1-scores of 96.42\% and 95.82\%, respectively, demonstrating strong performance across different CMS sizes. These results validate the effectiveness of our sketch-based syscall monitoring and classification approach. In addition, we found that lightweight machine learning classifiers such as Random Forest (RF) can serve as efficient alternatives to deep learning models, achieving an average F1-score of 95.5\%.

\noindent\textbf{Effectiveness of Phase-2 Classification.} We evaluated the ability of our classifiers to distinguish cryptojacking processes or containers from their benign counterparts. This multi-class classification problem (Section~\ref{sec:detection}) involved three categories: (i)~XMRig, (ii)~Sysrv, and (iii)~Benign. The first two categories represent the cryptojacker families detailed in Table~\ref{table:cryptojacker_family}, while the third category includes benign processes. The classifier's goal is to accurately classify a given process or container into one of these categories. To determine the efficacy of our sliding window-based monitoring in detecting cryptojackers, we tested various sliding window sizes: $\Delta T=30$, $60$, and $90$ seconds (Section~\ref{sec:monitoring}), with larger windows allowing for the analysis of more syscall sequences.

Table~\ref{table:phase2_results} summarizes the results of the Phase-2 classification. Our LSTM and CNN classifiers outperformed the RF classifier, achieving average F1-scores of 87.86\% and 96.67\%, respectively. The RF and LSTM classifiers struggled to accurately classify XMRig families due to their similar syscall patterns, whereas the CNN classifier excelled by capturing short-term temporal dependencies. Notably, for $\Delta T = 30$, the CNN classifier achieved F1-scores of 95.62\%, 98.54\%, and 98.87\% for XMRig, Sysrv, and Benign classes, respectively, underscoring its effectiveness in real-time detection scenarios. These results indicate that even a small sliding window is sufficient for achieving high accuracy, demonstrating the efficiency of \ourtool{}'s lightweight monitoring approach. While RF---one of the strongest traditional ML models---performed comparably in Phase-1, it failed to keep up in Phase-2. This performance gap highlights the limitations of classical models in capturing the obfuscated and fine-grained syscall behaviors exhibited by stealthy cryptojackers, which are better handled by deep learning models.

\noindent
\textbf{Classifying Cryptojackers and Non-cryptojackers.}
One concern is that \ourtool{}'s models may incorrectly classify general malware as cryptojackers. To evaluate this, we extended our experiment by incorporating real-world Mirai~\cite{antonakakis2017understanding} samples, which represent non-cryptojacking malware, and generated a total of 1,156 data instances through dynamic execution. This choice was motivated by the fact that Sysrv~\cite{sysrv}, one of the cryptojackers in our dataset, is also botnet-based and may therefore exhibit behavioral similarities to Mirai. Table~\ref{table:presumed} presents the evaluation results of the extended \ourtool{} classifier. With a sliding window size of $\Delta T = 30$, the CNN-based classifier achieved F1-scores of 98.46\%, 95.09\%, 97.69\%, and 99.02\% for the Mirai, XMRig, Sysrv, and Benign classes, respectively. These results demonstrate that \ourtool{} can effectively distinguish cryptojackers from unrelated malware families such as Mirai, even in the presence of behavioral overlap.

\begin{table}[t]
\caption{Extended Phase-2 classification results of \ourtool{} with additional non-cryptojacking malware (Mirai) to evaluate discrimination capability.}
\vspace{-0.1in}
\label{table:presumed}
\centering
\resizebox{\linewidth}{!}{
\begin{tabular}{cc ccc ccc}
\toprule

    \multirow{3}{*}{\textbf{Sliding Window Size}} & \multirow{3}{*}{\textbf{Class}} & \multicolumn{3}{c}{\textbf{LSTM}} & \multicolumn{3}{c}{\textbf{CNN}} \\ \cmidrule(lr){3-5} \cmidrule(lr){6-8}
    & & Precision & Recall & F1-score & Precision & Recall & F1-score \\ 
    \midrule

    \multirow{5}{*}{\makecell{$\Delta T=30$}} 
    & Mirai & 0.8168 & 0.9512 & 0.8789 & 0.9915 & 0.9778 & 0.9846\\
    & XMRig & 0.8125 & 0.6887 & 0.7455 & 0.9443 & 0.9576 & 0.9509\\
    & Sysrv  & 0.9892 & 0.9109 & 0.9485 & 0.9769 & 0.9769 & 0.9769\\ 
    & Benign & 0.9419 & 0.9643 & 0.9530 & 0.9895 & 0.9910 & 0.9902\\  \cmidrule(lr){3-5} \cmidrule(lr){6-8}
    & Average & 0.8901 & 0.8788 & 0.8815 & \textbf{0.9755} & \textbf{0.9758} & \textbf{0.9757} \\ \midrule

    \multirow{5}{*}{\makecell{$\Delta T=60$}}
    & Mirai & 0.9067 & 0.9645 & 0.9347 & 0.9793 & 0.9595 & 0.9693\\
    & XMRig & 0.7151 & 0.9110 & 0.8012 & 0.9018 & 0.9363 & 0.9187\\
    & Sysrv & 0.9648 & 0.8253 & 0.8896 & 0.9459 & 0.9211 & 0.9333\\ 
    & Benign & 0.9858 & 0.9206 & 0.9521 & 0.9893 & 0.9920 & 0.9907\\ \cmidrule(lr){3-5} \cmidrule(lr){6-8}
    & Average & 0.8931 & 0.9054 & 0.8944 & \textbf{0.9541} & \textbf{0.9522} & \textbf{0.9530} \\ \midrule

    \multirow{5}{*}{\makecell{$\Delta T=90$}} 
    & Mirai & 0.8280 & 0.9673 & 0.8922 & 0.9628 & 0.9673 & 0.9650\\
    & XMRig & 0.7209 & 0.5345 & 0.6139 & 0.8966 & 0.9163 & 0.9442\\
    & Sysrv & 0.9521 & 0.8910 & 0.9205 & 0.9317 & 0.9615 & 0.9464\\ 
    & Benign & 0.9377 & 0.9517 & 0.9446 & 0.9888 & 0.9851 & 0.9870\\ \cmidrule(lr){3-5} \cmidrule(lr){6-8}
    & Average & 0.8597 & 0.8361 & 0.8428 & \textbf{0.9551} & \textbf{0.9526} & \textbf{0.9537} \\ \midrule
    
    \multirow{4}{*}{Average} & Mirai & 0.8505 & 0.9610 & 0.9019 & 0.9778 & 0.9682 & 0.9729 \\ & XMRig & 0.7495 & 0.7114 & 0.7202 & 0.9142 & 0.9367 & 0.9379 \\
    & Sysrv & 0.9687 & 0.8757 & 0.9195 & 0.9515 & 0.9531 & 0.9552 \\
    & Benign & 0.9551 & 0.9455 & 0.9499 & 0.9892 & 0.9893 & 0.9893 \\
    
    \bottomrule
\end{tabular}
}
\vspace{-0.2in}
\end{table}

\begin{table}[t]
\caption{Comparison of the detection performance of three existing solutions~\cite{tanana2020wmi,darabian2020syscall,sanda2023barbhuiya} with \ourtool{}.}
\vspace{-0.1in}
\label{table:mock}
\centering
\resizebox{\linewidth}{!}{
\begin{tabular}{c c c c c c c c}
\toprule

    \multirow{4}{*}{\makecell{\textbf{Sliding}\\\textbf{Window}\\\textbf{Size}}} & \multirow{4}{*}{\textbf{Class}} &\multirow{2}{*}{\textbf{Tanana \emph{et al.}~\cite{tanana2020wmi}}} & \textbf{Sanda \emph{et al.}~\cite{sanda2023barbhuiya}} &\multicolumn{2}{c}{\textbf{Darabian \emph{et al.}~\cite{darabian2020syscall}}} &\multicolumn{2}{c}{\textbf{\ourtool{}}} \\ \cmidrule(lr){4-4} \cmidrule(lr){5-6} \cmidrule(lr){7-8}
    
    &   & &\multicolumn{1}{c}{\textbf{MLP}} &\multicolumn{1}{c}{\textbf{LSTM}} & \multicolumn{1}{c}{\textbf{CNN}} & \multicolumn{1}{c}{\textbf{LSTM}} & \multicolumn{1}{c}{\textbf{CNN}} \\ 
    \cmidrule(lr){3-3} \cmidrule(lr){4-4} \cmidrule(lr){5-5} \cmidrule(lr){6-6} \cmidrule(lr){7-7} \cmidrule(lr){8-8}
    
    & & TPR & TPR & TPR & TPR & TPR & TPR \\ 
    \midrule

    \multirow{2}{*}{\makecell{$\Delta T=30$}} 
    & XMRig & 0.6496 & 0.7869 & 0.8428 & 0.9142 & 0.9084 & \textbf{0.9424}\\
    & Sysrv & 0.6420 & 0.8765 & 0.7175 & 0.9707 & 0.9090 & \textbf{0.9837}\\ 
    \midrule

    \multirow{2}{*}{\makecell{$\Delta T=60$}}
    & XMRig & 0.7476 & 0.7372 & 0.7731 & 0.8969& 0.8762 & \textbf{0.9072}\\
    & Sysrv & 0.6742 & 0.8446 & 0.7419 & 0.9612 & 0.8967 & \textbf{0.9741}\\ 
    \midrule

    \multirow{2}{*}{\makecell{$\Delta T=90$}} 
    & XMRig & 0.8120 & 0.7333 & 0.7291 & 0.8666 & 0.7647 & \textbf{0.9207}\\
    & Sysrv & 0.7125 & 0.8684 & 0.7172 & 0.9509 & 0.6604 & \textbf{0.9801}\\ 
    \midrule
    
    \multirow{2}{*}{Average} & XMRig & 0.7364 & 0.7524 & 0.7816 & 0.8925 & 0.8497 & \textbf{0.9234}\\
    & Sysrv & 0.6762 & 0.8631 & 0.7255 & 0.9609 & 0.8220 & \textbf{0.9793}\\
    \bottomrule
\end{tabular}
}
\vspace{-0.22in}
\end{table}

\vspace{-0.1in}
\subsection{Comparison with Existing Solutions}

To evaluate whether \ourtool{} outperforms existing host-based cryptojacking detection methods, we conducted comparative experiments with three state-of-the-art approaches~\cite{tanana2020wmi,darabian2020syscall,sanda2023barbhuiya}. Tanana \emph{et al.}~\cite{tanana2020wmi} proposed a threshold-based method that flags processes exceeding a CPU usage of 10\%. Sanda~\emph{et al.}~\cite{sanda2023barbhuiya} presented a deep learning-based detector that leverages CPU usage patterns, while Darabian~\emph{et al.}~\cite{darabian2020syscall} introduced deep learning models designed to learn syscall sequences using multiple architectures. Since the source code for these methods is not publicly available, we re-implemented each approach based on the algorithmic details provided in their papers. Specifically, we applied the 10\% CPU threshold for Tanana~\emph{et al.}, and replicated the MLP (Multi-Layer Perception), LSTM, and CNN models described by Sanda~\emph{et al.} and Darabian~\emph{et al.} It is worth noting that the Count-Min Sketch (CMS)-based Phase-1 classification used in \ourtool{} is not applicable to these prior works; therefore, we restricted our evaluation to Phase-2 classification using sliding windows over CPU usage or syscall sequences, depending on each method’s original design.

Table~\ref{table:mock} presents the results of these experiments.
The best performance reported by Tanana~\emph{et al.} was a True Positive Rate (TPR) of 81.2\% on the XMRig family when using a sliding window size of \( \Delta T = 90 \).
Similarly, Sanda \emph{et al.} achieved a TPR of 87.65\% when tested on the Sysrv family using a sliding window size of \( \Delta T = 30 \). These lower TPR are attributed to the presence of CPU-throttled cryptojackers with reduced CPU activity, highlighting the limitations of CPU-based thresholds in detecting stealthy cryptojackers, as discussed in Section~\ref{sec:motivation}. Darabian \emph{et al.} outperformed Tanana \emph{et al.}, achieving mean TPRs of 75.35\% and 92.67\% for their LSTM and CNN models, respectively. However, these results still fall short compared to \ourtool{}. Additionally, their solution suffers from scalability challenges due to inefficiencies in syscall monitoring. In contrast, \ourtool{} leverages sketch- and sliding window-based monitoring techniques, enabling efficient and scalable detection of host-based cryptojackers in cloud environments. For a comparative analysis of detection performance, we also provide the ROC curves of Tanana \emph{et al.} and \ourtool{} in Appendix~\ref{append:roc}.

\begin{figure}[t]
\centering
\begin{minipage}[t]{0.48\linewidth}
    \centering
    \includegraphics[width=\linewidth,trim={0.5cm 0.5cm 0.3cm 0.2cm},clip]{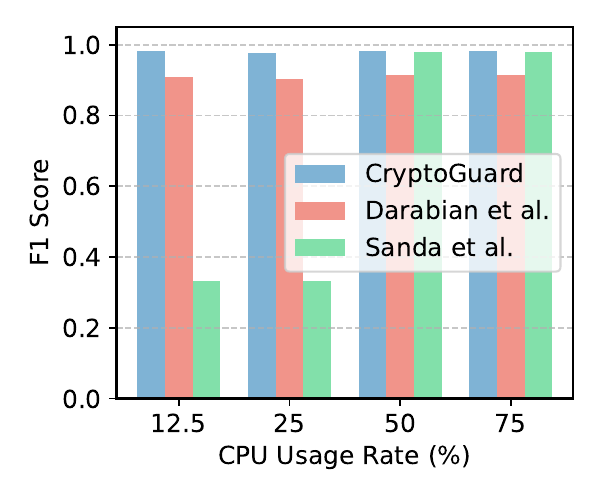}
    \caption{F1 scores at different CPU usage rates.}
    \label{fig:evasion}
\end{minipage}
\hfill
\begin{minipage}[t]{0.48\linewidth}
    \centering
    \includegraphics[width=\linewidth,trim={0.5cm 0.5cm 0.3cm 0.2cm},clip]{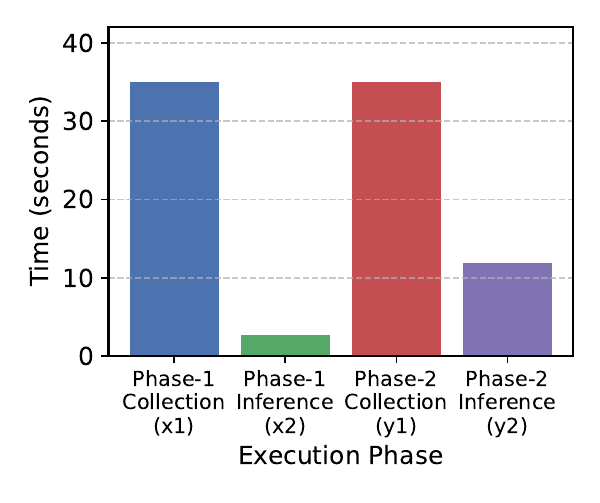}
    \caption{Latency of Phase-1 and Phase-2 collection and inference steps.}
    \label{fig:latency}
\end{minipage}
\vspace{-0.25in}
\end{figure}

\noindent
\textbf{Robustness to Evasion Attacks.}
As discussed in Section~\ref{sec:motivation}, host-based cryptojackers often employ CPU throttling to reduce resource usage and mimic benign behavior. We assessed the robustness of \ourtool{} against this evasion strategy and compared it with two prior methods: the syscall-based approach by Darabian~\emph{et al.}~\cite{darabian2020syscall} and a deep learning-based method by Sanda~\emph{et al.}~\cite{sanda2023barbhuiya} that leverages CPU usage patterns. Figure~\ref{fig:evasion} presents the F1-scores of the three approaches under different CPU useage rates of 12.5\%, 25\%, 50\%, and 75\%. Darabian~\emph{et al.} achieved F1-scores of 90.94\%, 90.23\%, 91.46\%, and 91.40\%, while Sanda~\emph{et al.} performed well at higher rates (98.00\% at 50\% and 75\%) but poorly at lower rates (33.33\% at 12.5\% and 25\%). In contrast, \ourtool{} consistently maintained high performance across all levels, achieving 98.15\%, 97.67\%, 98.34\%, and 98.13\%, respectively. These results indicate that our syscall-based detection offers greater resilience to CPU throttling-based evasion while maintaining lightweight efficiency, whereas CPU-based approaches may falter when resource usage is intentionally reduced.

\begin{figure}[t]
\centering
\begin{minipage}[t]{0.48\linewidth}
    \centering
    \includegraphics[width=\linewidth,trim={0.4cm 0.4cm 0.3cm 0.1cm},clip]{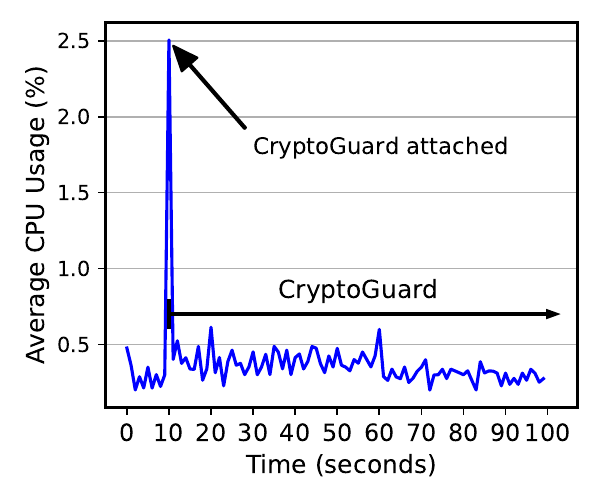}
    \caption{CPU usage variation.}
    \label{fig:cpu_variation}
\end{minipage}    
\hfill
\begin{minipage}[t]{0.48\linewidth}
    \centering
    \includegraphics[width=\linewidth,trim={0.4cm 0.4cm 0.3cm 0.1cm},clip]{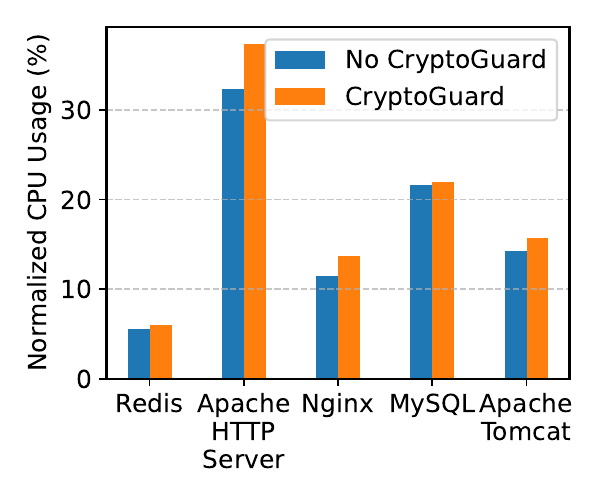}
    \caption{CPU usages of benign applications under workloads.}
    \label{fig:cpu_workload}
\end{minipage}
\vspace{-0.25in}
\end{figure}

\subsection{System Overhead}

We conducted experiments to measure the overhead of \ourtool{} on a single host to evaluate its viability as a lightweight solution for resource-limited cloud environments.

\noindent\textbf{CPU Overhead.} We assessed the CPU overhead of \ourtool{} compared to the baseline scenario where it is not deployed. Measuring overhead during the collection of behavioral features on the end-user side is crucial for conducting dynamic ML-based analysis~\cite{tekiner2021sok}, and \ourtool{} meets this standard. Figure~\ref{fig:cpu_variation} illustrates the CPU variation of \ourtool{} when it performs syscall monitoring. The baseline CPU usage was 0.29\% in the initial 10 seconds. During the syscall monitoring phase using eBPF, the average CPU usage increased to 0.35\%. The peak usage occurred when \ourtool{} was attaching but decreased quickly. This demonstrates that \ourtool{} imposes minimal overhead, avoiding interruptions to other benign processes.

\noindent\textbf{Latency Overhead.}
We measured the latency of each step in \ourtool{}, from Phase-1 collection and inference to Phase-2 processing. Figure~\ref{fig:latency} presents the average execution times over 100 test cases, using a Count-Min Sketch of size $55 \times 5$ and a sliding window size of $\Delta T = 30$. In Phase-1, the collection time (x1) was approximately 35 seconds, and the inference time (x2) was 2.68 seconds. In Phase-2, the collection time (y1) was also approximately 35 seconds, and the inference time (y2) was 11.87 seconds. The total end-to-end latency for executing the detection pipeline is approximately 85 seconds, which is reasonable given that cryptojackers must engage in long-term mining to generate profit.

\noindent
\textbf{Overhead on Benign Applications.}  
We evaluated whether \ourtool{} introduces significant overhead on benign applications, given that it monitors the syscalls of all processes using eBPF. To conduct this experiment, we configured two virtual machines: one VM was used to generate workload using appropriate benchmarking tools, while the other hosted a benign server application to process incoming requests. The benchmarking tools were chosen based on the type of server: \texttt{ab} was used for Apache HTTP Server, Nginx, and Tomcat; \texttt{redis-benchmark} for Redis; and \texttt{mysqlslap} for MySQL. We measured the CPU usage of each server application under two configurations: (i) without \ourtool{}, and (ii) with \ourtool{} enabled. As shown in Figure~\ref{fig:cpu_workload}, the CPU usage increased by an average of only 10.97\% when \ourtool{} was active. These results demonstrate that \ourtool{} imposes minimal overhead on benign applications.

\begin{figure}[t]
\subfloat[Creation time.]{
        \includegraphics[width=0.47\linewidth,trim={0.3cm 0.5cm 0.3cm 0.3cm},clip]{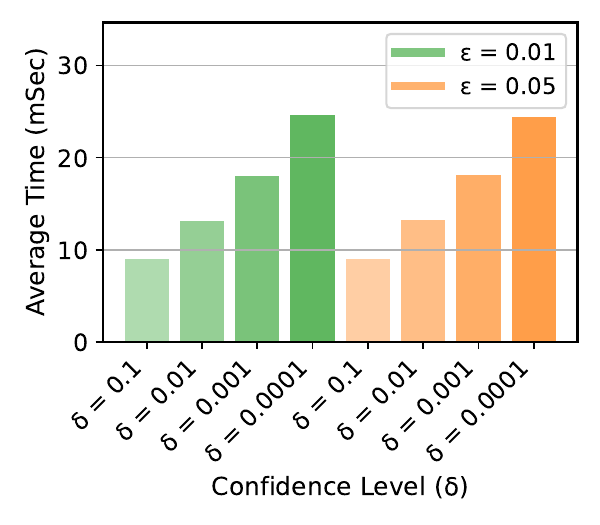}
        \label{fig:cms_time}
}\hfill
\subfloat[Memory usage.]{
        \includegraphics[width=0.47\linewidth,trim={0.3cm 0.4cm 0.3cm 0cm},clip]{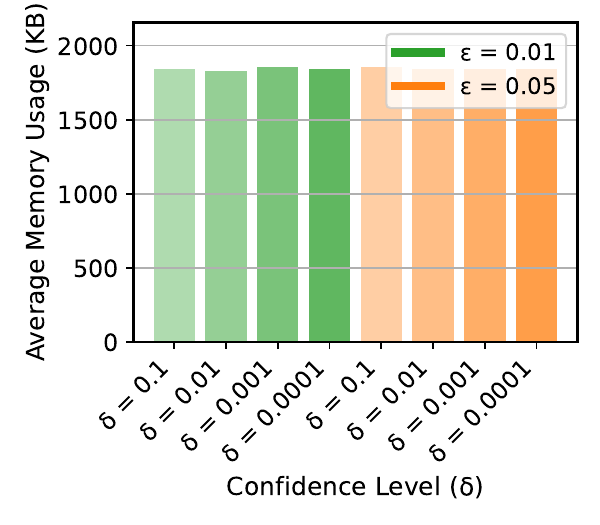}
        \label{fig:cms_memory}
}
\vspace{-0.1in}
\caption{Creation time and memory overhead for Count-Min Sketches (CMS) of varying sizes determined by the error rate ($\epsilon$) and confidence level ($\delta$).}
\vspace{-0.2in}
\end{figure}

\noindent\textbf{Sketch Overhead.} We evaluated the overhead of CMS to determine their suitability for cloud environments by measuring the creation time and memory usage for various CMS configurations based on the error rate ($\epsilon$) and confidence level ($\delta$). Figure~\ref{fig:cms_time} shows the creation times for different CMS parameters. The results indicate that creation time increases with higher confidence levels, suggesting that a more accurate and larger CMS requires additional time to construct. Additionally, variations in the error rate did not affect the creation time. Notably, the peak creation time observed was a maximum of 24ms, demonstrating the efficiency of CMS creation. Figure~\ref{fig:cms_memory} illustrates the memory usage of \ourtool{}'s Agent during CMS creation. The Agent exhibited an average memory usage of 1,840KB, indicating a minimal footprint, independent of the sketch parameters.

\section{Limitations}
\label{sec:discussion}

This section discusses the limitations of \ourtool{}. Note that additional discussions are provided in Appendix~\ref{appen:discussion}.

\noindent\textbf{Detecting Advanced Evasion Attacks.}
While we demonstrate that \ourtool{} is capable of detecting cryptojackers based on their syscall patterns for both typical and obfuscated behaviors, more advanced evasion techniques beyond CPU throttling and PID obfuscation could circumvent our methodology. Specifically, attackers may adopt three evasion strategies. First, cryptojackers often spawn child processes as part of their obfuscation strategy. For example, a dropper may repeatedly create new processes to download and execute mining software, thereby complicating detection efforts. Second, cryptojackers may distribute their syscall invocations across different processes~\cite{botacin2019vanilla} to evade per-process monitoring. While our sketch-based monitoring remains robust against this attack due to its host-level aggregation, the sliding window-based per-process monitoring is susceptible to evasion. Third, cryptojackers could interleave benign-like behavior among their malicious activities to mimic benign processes. Addressing this would require longer-term monitoring to correlate distributed evidence across processes and time. We leave the mitigation of these advanced evasion techniques as future work.

\noindent
\textbf{Risks of Process Name Spoofing.}
\ourtool{} detects cryptojackers by tracking process name–PID pairs, which remains effective even when attackers obfuscate their PIDs. To evade this methodology, advanced cryptojackers might spoof their process names to match those of benign processes, potentially leading to false termination. However, \ourtool{} minimizes such risks by relying on multiple lines of evidence beyond names alone. Specifically, it incorporates syscall behavior patterns, persistent PID tracking, and detection of poisoned startup entry points (e.g., \texttt{rc.local}, \texttt{crontab}) before making termination decisions. These combined indicators are highly characteristic of cryptojackers, significantly reducing the likelihood of false positives. Furthermore, widely used benign processes can be explicitly whitelisted to avoid accidental termination.

\vspace{-0.05in}
\section{Related Work}
\label{subsec:related_work}

\noindent\textbf{Cryptojacking Landscape.} Cryptojackers have become a significant threat, driven by the emergence of new technologies and infrastructures that adversaries can exploit. For example, WebAssembly (Wasm), a portable executable format that runs in web browsers, has enabled the creation of Wasm-based in-browser cryptojackers~\cite{tekiner2021sok}. Adversaries also employ advanced techniques, such as stealth mining pools~\cite{zhang2023under} and binary obfuscation~\cite{bhansali2022first}, to obscure their cryptojacking activities. Alarmingly, Hong \emph{et al.}~\cite{hong2018you} estimated that cryptojackers could generate approximately 59,000 USD in profits. More recently, the rise of host-based cryptojackers, which exploit virtually unlimited cloud resources, presents an even greater threat. As discussed in this paper, detecting and mitigating such attacks is a timely and challenging topic.

\noindent\textbf{Cryptojacking Detection.} In addition to the studies on host-based cryptojackers discussed in Section~\ref{subsec:existing_work}~\cite{barbhuiya2018rads,ahmad2020dca,mani2020decryptopro,gomes2020cryingjackpot,gangwal2020hpc,darabian2020syscall,lachtar2020rsx,tanana2020wmi,carprolu2021noise,tekiner2022ali,sanda2023barbhuiya,almurshid2023binary,franco2023honeypot}, various solutions have been proposed to counter in-browser cryptojackers. For example, MINOS~\cite{naseem2021minos} models Wasm binaries as grayscale images and utilizes Convolutional Neural Networks (CNNs) to detect cryptojacking activity. MineThrottle~\cite{bian2020minethrottle} leverages block-level profiling to measure CPU usage within browsers, effectively identifying cryptojackers. Similarly, MagTracer \cite{xiao2023magtracer} targets GPU-based cryptojackers by analyzing magnetic leakage signals emitted by GPUs. In contrast, our work is dedicated to the effective and efficient detection of host-based cryptojackers in cloud environments, coupled with practical remediation strategies.

\vspace{-0.05in}
\section{Conclusion}

Detecting and remediating host-based cryptojacking threats is a significant challenge due to their stealthy behavior and the vast number of hosts and processes in cloud environments. This paper introduces \ourtool{}, a lightweight hybrid solution for addressing this issue. We propose innovative sketch-based and sliding window-based monitoring techniques to efficiently trace syscall patterns of cryptojackers. Our evaluation demonstrates that these methods are effective in detecting cryptojackers using partial information. Additionally, we develop remediation strategies that target the persistence mechanisms employed by cryptojackers. We believe our research offers a new direction for combating host-based cryptojackers, contributing to safer cloud environments.

\vspace{-0.05in}
\begin{acks}
We would like to thank the anonymous reviewers for their valuable comments and the shepherd for their thorough guidance during the revision process. This work was partly supported by the National Research Foundation of Korea (NRF) grant funded by the Korea government (MSIT) (No. RS-2024-00457937, Design and implementation of security layers for secure WebAssembly-based serverless environments, 50\%) and the Institute of Information \& Communications Technology Planning \& Evaluation (IITP) grant funded by the Korea government (MSIT) (No. 2021-0-00118, RS-2021-II210118, Development of decentralized consensus composition technology for large-scale nodes, 50\%).
\end{acks}

\bibliographystyle{ACM-Reference-Format}
\bibliography{reference.bib}

\appendix

\section{Deep Learning Model Architecture}
\label{appen:model_arch}

\begin{figure}[h!]
    \centering
    \subfloat[CNN]{
        \includegraphics[width=.9\linewidth]{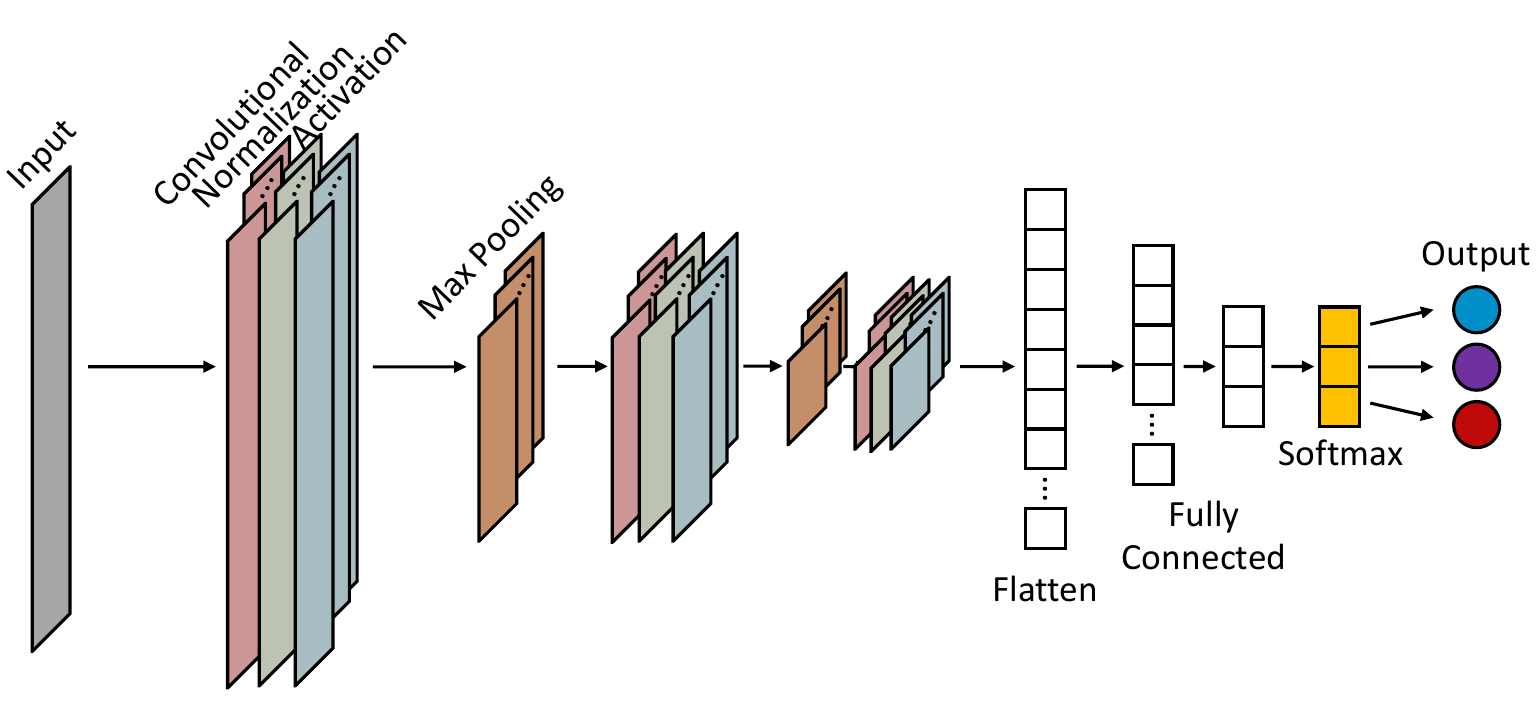}
    }
    
    \subfloat[LSTM]{
        \includegraphics[width=.9\linewidth]{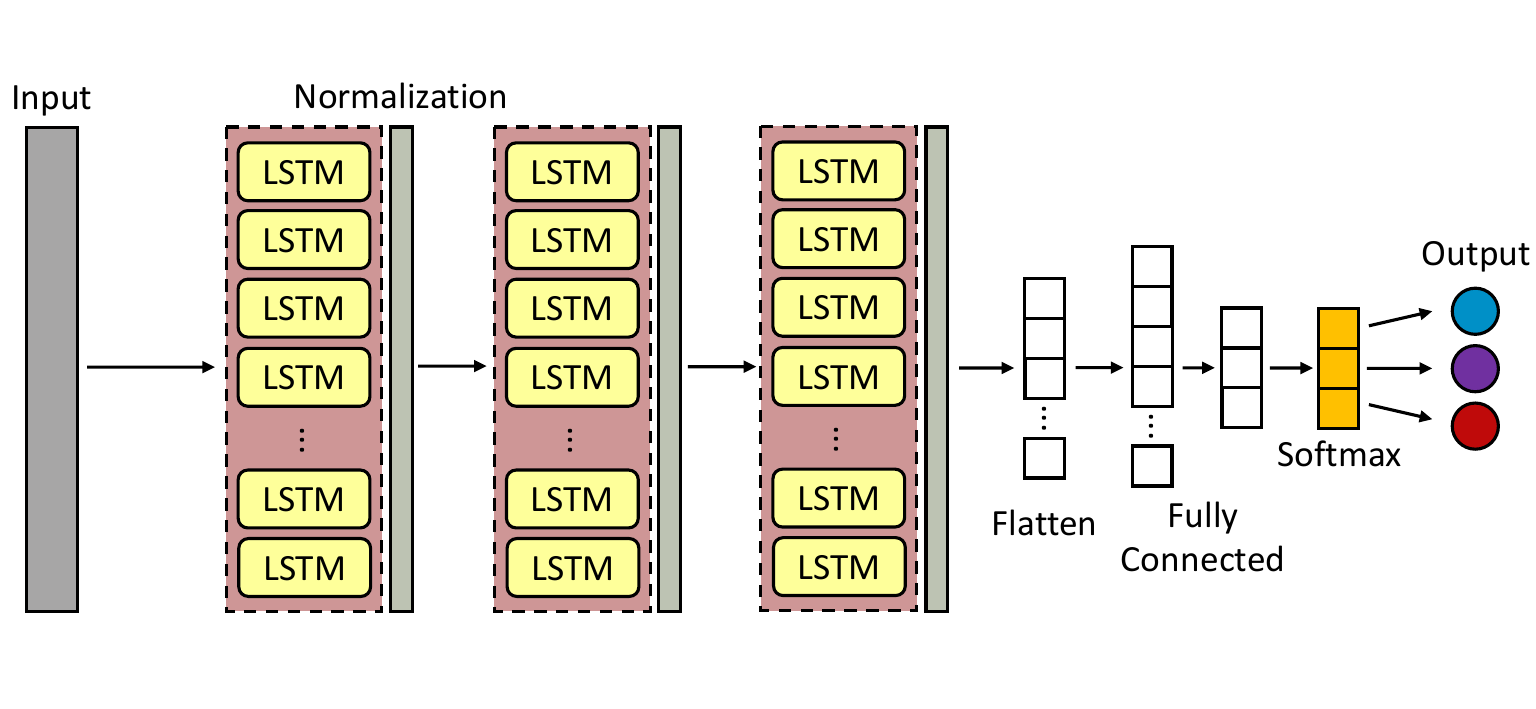}
    }
    \caption{The architecture of our deep learning models.}
    \label{fig:model_architecture}
    \vspace{-0.1in}
\end{figure}

\section{Remediation APIs and a Use Case}
\label{appen:algorithm}

Remediation strategies often vary based on the specific needs of administrators. To address these diverse requirements, \ourtool{} provides a set of APIs that facilitate tailored remediation strategies, as summarized in Table~\ref{tab:prevention_api}. For instance, a private cloud administrator may prioritize extended monitoring and behavioral analysis of a cryptojacker while minimizing its impact through quarantine, whereas a public cloud tenant may focus on immediate removal to prevent excessive costs. Algorithm~\ref{algo1} in Appendix~\ref{appen:algorithm} exemplifies the former scenario, where \ourtool{}'s Remediator periodically scans processes on a suspicious host to identify cryptojackers (lines 1-3). Upon detection, the Remediator blocks the process’s network connections to disrupt communication with mining pools (line 4) and continues monitoring its network traffic and system calls over a specified duration \( T \) (lines 5-8).

\renewcommand\cellalign{tl}  

\begin{table}[!t]
    \centering
    \caption{APIs provided by \ourtool{} for cryptojacking remediation.}
    \scriptsize
    \begin{tabular}{
        >{\raggedright\arraybackslash}p{3.5cm}
        >{\raggedright\arraybackslash}p{4.0cm}
    }
        \toprule
        \textbf{API} & \textbf{Description} \\
        \midrule

        \texttt{isCryptojacker(pid)} &
        \makecell[l]{Detects whether a process (\texttt{pid})\\ is a cryptojacker.} \\ \midrule

        \texttt{blockNetwork(pid, dstIp)} &
        \makecell[l]{Blocks all network connections of a process\\ (\texttt{pid}) to the specified destination IP\\ (\texttt{dstIp}).} \\ \midrule

        \texttt{traceNetwork(pid, T)} &
        \makecell[l]{Monitors and logs the network traffic\\ generated by a process (\texttt{pid}) for a\\ period \texttt{T}.} \\ \midrule

        \texttt{blockSyscalls(pid, syscalls)} &
        \makecell[l]{Restricts a process (\texttt{pid}) from executing\\ a specific set of system calls (\texttt{syscalls}).} \\ \midrule

        \texttt{traceSyscalls(pid, T)} &
        \makecell[l]{Tracks and logs the system calls executed\\ by a process (\texttt{pid}) for a period \texttt{T}.} \\ \midrule

        \texttt{terminateProcess(pid, pname)} &
        \makecell[l]{Sends a \texttt{SIGKILL} signal to immediately\\ terminate a process (\texttt{pid}) identified by\\ its name (\texttt{pname}).} \\ \midrule

        \texttt{deleteProcessEntry(pid, pname, entrypoint)} &
        \makecell[l]{Finds a process (\texttt{pid}) with the name\\ (\texttt{pname}) and entry point (\texttt{entrypoint}),\\ and removes its entry.} \\ \midrule

        \texttt{deleteBinary(pid, path)} &
        \makecell[l]{Identifies a process (\texttt{pid}) using its\\ binary path (\texttt{path}) and deletes the binary.} \\

        \bottomrule
    \end{tabular}
    \label{tab:prevention_api}
\end{table}

\begin{algorithm}[h!]
\footnotesize
\SetAlgoLined
\KwIn{Set of suspicious processes $P$, monitoring period $T$}
\KwOut{Behavior logs for the cryptojacker process}
\ForEach{$pid \in P$}{
    \If{\texttt{isCryptojacker(pid)}}{
        \texttt{dstIp} $\gets$ \texttt{getNetworkAddr(pid)}\;

        \texttt{blockNetwork(pid, dstIp)}\;
    
        \texttt{traceNetwork(pid, T)}\;
        \texttt{log("Network traffic traced for process $pid$")}\;

        \texttt{traceSyscalls(pid, T)}\;
        \texttt{log("System calls traced for process $pid$)}\;
    }
}
\caption{An example of long-term monitoring}
\label{algo1}

\end{algorithm}

\section{ROC Curves of Deep Learning Models}
\label{append:roc}

\begin{figure}[t]
\subfloat[LSTM ROC curves.]{
        \includegraphics[width=\linewidth,trim={0.3cm 0cm 0.3cm 0.3cm},clip]{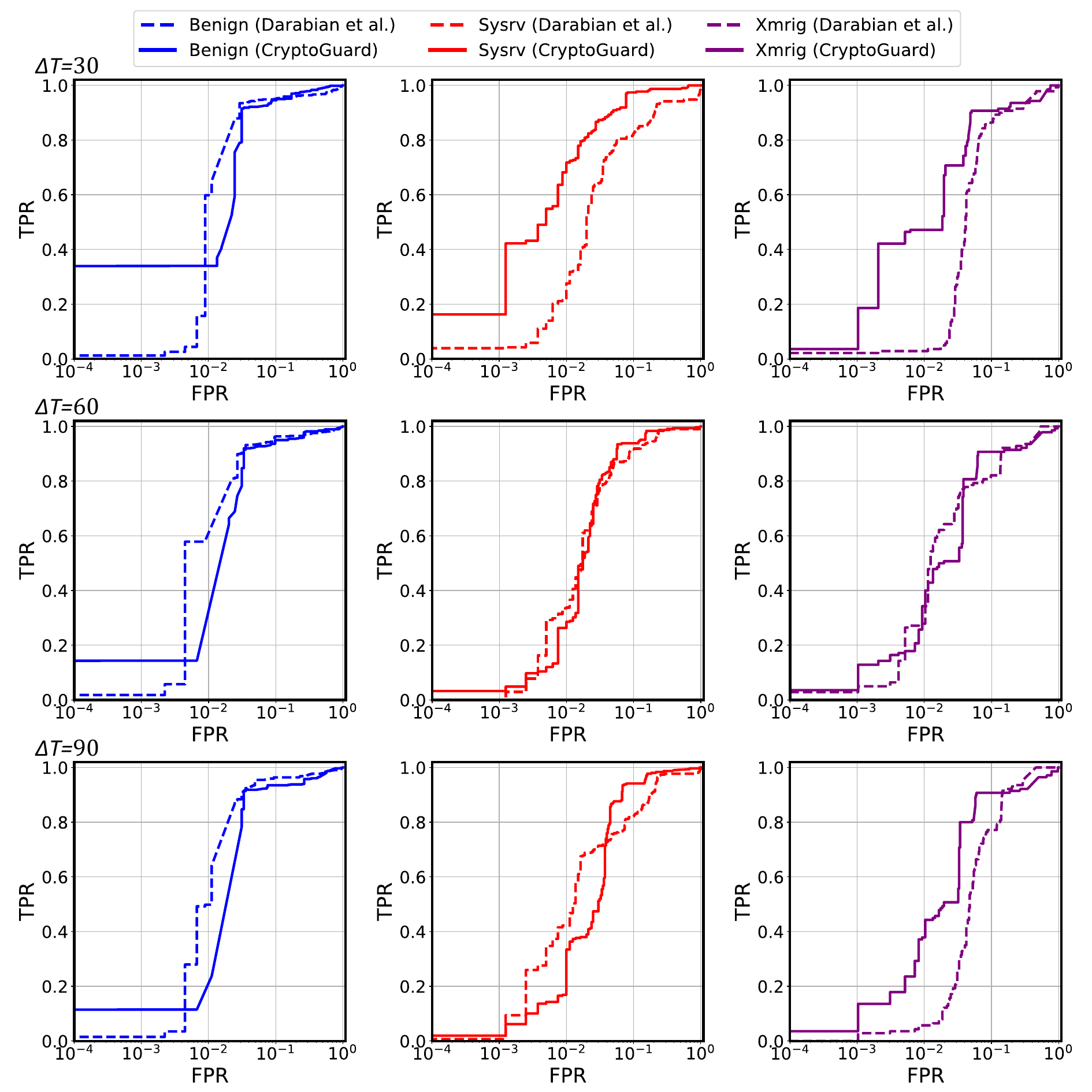}
        \label{fig:roc_lstm}
}

\subfloat[CNN ROC curves.]{
        \includegraphics[width=\linewidth,trim={0.3cm 0cm 0.3cm 0cm},clip]{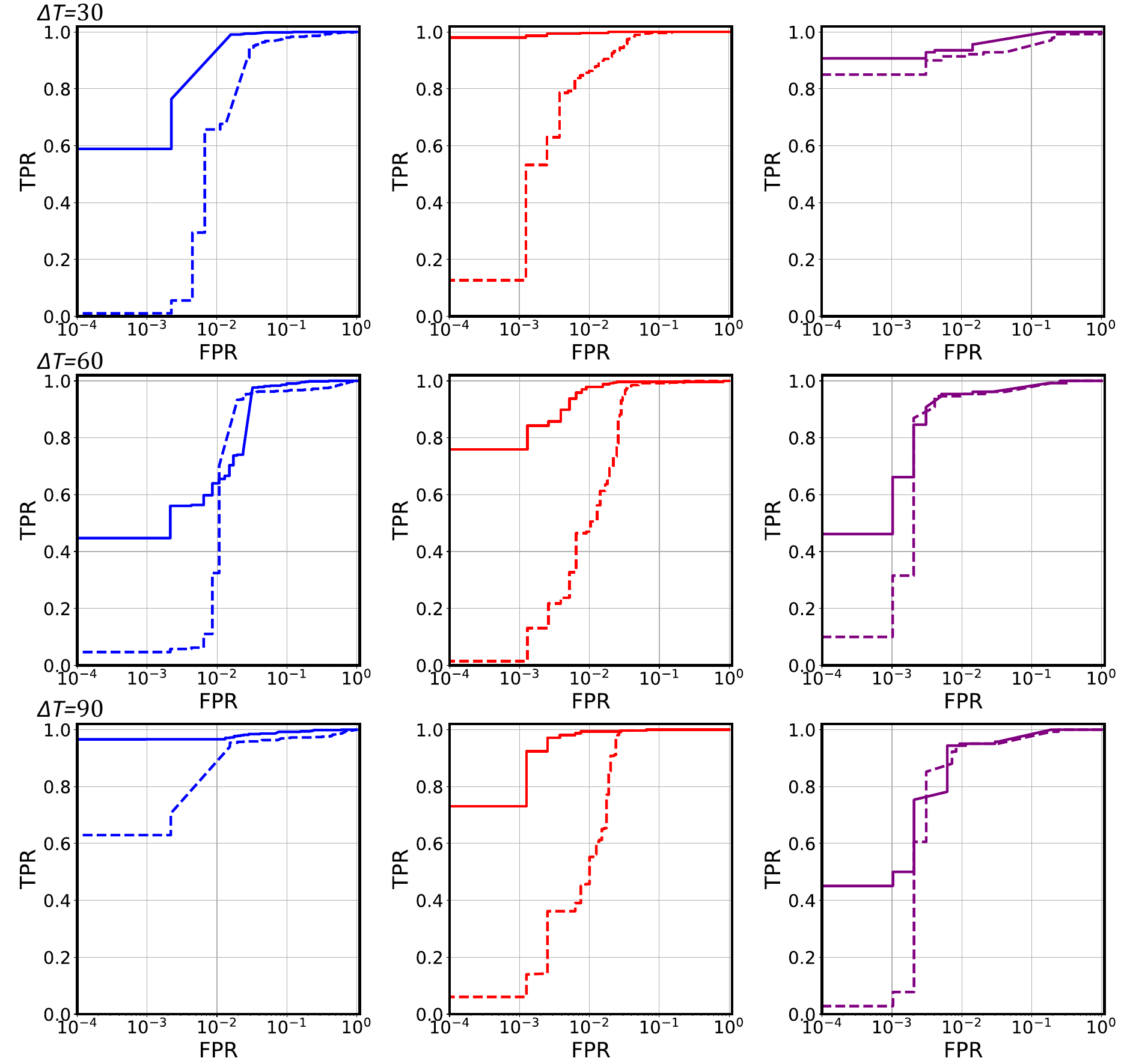}
        \label{fig:roc_cnn}
}
\caption{ROC curves of the existing solution~\cite{darabian2020syscall} (dashed lines) and \ourtool{} (solid lines) for LSTM and CNN models across different sliding window sizes ($\Delta T=30, 60, 90$).}
\label{fig:roc_curve_append}
\vspace{-0.1in}
\end{figure}

To compare the performance of \ourtool{} with that of Darabian~\emph{et al.}, Figure~\ref{fig:roc_curve_append} presents the ROC curves for deep learning-based solutions across each class (i.e., Benign, Sysrv, XMRig) and for different sliding windows (\( \Delta T = 30, 60, 90 \)). The TPR (True Positive Rate) for the Sysrv and XMRig classes indicates the proportion of benign processes incorrectly classified as cryptojackers. Therefore, achieving high TPR values at low FPRs (False Positive Rates) is critical for ensuring system reliability during cryptojacker detection and mitigation. The ROC curves for Darabian~\emph{et al.} show a substantial drop in TPR when the FPR is constrained to $\leq 10^{-2}$. In contrast, \ourtool{} maintains consistently high TPRs even at low FPRs, demonstrating its effectiveness in minimizing disruptions to benign processes while accurately identifying cryptojackers.

\section{Discussion}
\label{appen:discussion}

\noindent\textbf{Compatibility of Remediation with Containers.} The current persistence logic of \ourtool{} Remediator primarily focuses on VM and bare-metal hosts, as many cryptojacker samples in our experiments exploit entrypoints specific to these environments. However, to address the evolving logic of cryptojackers targeting containerized environments, particularly Kubernetes misconfigurations such as DaemonSets~\cite{kubenetes_persistence}, our remediation mechanism must be adapted accordingly. These adjustments are expected to be minor, as the Remediator's Kubernetes implementation (Section~\ref{subsec:implementation}) can be extended to scan for malicious DaemonSets associated with cryptojackers during deployment. Since this approach requires collecting cryptojacker samples specific to Kubernetes environments, we leave this aspect for future work.

\noindent\textbf{Comparison with Anti-Virus Software.} Cryptojackers could potentially be detected and remediated using anti-virus (AV) software designed for Linux environments. While AV solutions can perform tasks similar to \ourtool{}, they have several limitations. First, most AV software relies on malware signature recognition, rendering it ineffective against unknown cryptojackers. In contrast, \ourtool{} can detect new variants by leveraging syscall patterns learned through deep learning models. Second, AV solutions often require substantial system resources, which is a significant drawback in resource-constrained cloud environments. By comparison, \ourtool{} is designed as a lightweight solution utilizing sketch-based syscall monitoring. Lastly, many AV tools are proprietary, whereas \ourtool{} is open and compatible with any Linux host supporting eBPF, a capability included in modern Linux kernels.

\end{document}